\documentclass{article}
\usepackage[margin=2.5cm]{geometry}
\usepackage{graphicx,subcaption,import}
\usepackage{amsmath,amssymb}
\usepackage{bbm}
\usepackage[utf8]{inputenc}
\usepackage{epstopdf, epsfig}
\usepackage[hidelinks]{hyperref}
\usepackage{authblk}
\usepackage{color}

\title{
Impact of deformation bands on fault-related fluid flow in field-scale simulations
}

\author[1]{Runar L. Berge \thanks{Corresponding author. E-mail address: runar.lie.berge@hivolda.no}}
\author[2,3]{Sarah E. Gasda}
\author[3]{Eirik Keilegavlen}
\author[2]{Tor Harald Sandve}
\affil[1]{Volda University College, Volda, Norway}
\affil[2]{NORCE Norwegian Research Centre, Bergen, Norway}
\affil[3]{University of Bergen, Bergen, Norway}
\date{March 2022}

\usepackage[backend=biber,
style=numeric-comp,
giveninits=true,
maxbibnames=24,
sorting=none,
url=false,
isbn=false,
doi=true,
natbib=true, 
hyperref=true, 
]{biblatex}
\bibliography{ref.bib}
\renewbibmacro{in:}{%
  \ifentrytype{article}
    {}
    {\bibstring{in}%
     \printunit{\intitlepunct}}}
\renewcommand\vec[0]{\ensuremath{\boldsymbol}} 

\usepackage{xspace}
\newcommand{\co}{CO\ensuremath{_\textrm{2}}\xspace}
\usepackage{setspace}
\usepackage{lineno}

\doublespace
\newif\ifpicture
\picturetrue

\begin{document}
\maketitle

\begin{abstract}
    Subsurface storage of \co is predicted to rise exponentially in response to the increasing levels of \co in the atmosphere. Large-scale \co injections into the subsurface require understanding of the potential for fluid flow through faults to mitigate risk of leakage. Here, we study how to obtain effective permeability of deformation bands in the damage zone of faults. Deformation bands are relatively small, low permeability features that can have a significant effect on flow dynamics, however, the discrepancy of scales is a challenge for field-scale simulation. A new analytical upscaling model is proposed in order to overcome some of the shortcomings of conventional upscaling approaches for heterogeneous porous media. The new model captures the fine-scale impact of deformation bands on fluid flow in the near-fault region, and can be derived from knowledge of large-scale fault properties. To test the accuracy of the model it is compared to fine-scale numerical simulations that explicitly include individual deformation bands. For a wide range of different stochastically generated deformation bands networks, the upscaling model shows improved estimate of effective permeability compared to conventional upscaling approaches. By applying the upscaling model to a full-field simulation of the Smeaheia storage site in the North Sea, we show that deformation bands with a permeability contrast higher than three orders of magnitude may act as an extra layer of protection from fluid flow through faults.
\end{abstract}


\section{Introduction}

\co storage is a growing industry with a clear role to play in reaching urgent climate targets \cite{IPCC_18}. Current projections for large-scale deployment of \co storage predict global injection rates reaching 20 Gt/y by 2100~\cite{Zahasky_Krevor_20}. Such rapid scale-up relies on developing a large number of storage sites in the coming decades, primarily along continental margins that are currently active oil and gas regions \cite{Ringrose_Meckel_19}. Global efforts to map \co storage resources indicate that many different types of sedimentary systems spanning a large range of depositional environments and paleotectonic settings may be suitable for \co storage (see e.g.~\cite{NPD_Atlas}). Identification and maturation of suitable \co storage sites relies on a number of geological, technical, economic, and socio-political factors~\cite{rothkirch2021,akhurst2021}. First-mover projects will naturally choose to develop optimal sites, but eventually many alternative prospective sites will need to be further matured and de-risked to meet growing storage demand.

Faulted sedimentary formations are promising options for \co storage given abundant evidence in petroleum exploration that fault seals can accumulate hydrocarbons over geologic timescales~\cite{hardman1991significance,roberts1990geometry}. Recent interest in assessing faulted reservoirs for \co storage~\cite{wu2021} underline the importance of understanding the role of faults with respect to \co containment and pressure communication within the storage reservoir and with its surroundings~\cite{yielding2011,bretan2011}. Although fault seal analysis is quite reliable in the hydrocarbon context (e.g., \cite{yielding2010,Knipe1997} and references therein), it is challenging to adapt these workflows to \co storage~\cite{karolyte2020}. The risk for along-fault fluid migration is based on natural analogues of \co leakage from faulted reservoirs \cite{shipton2004,dockrill2010}. Industrial experience in the In Salah CCS project also indicates that faults can be activated due to pressure build-up~\cite{rutqvist_coupled_2010,morris_study_2011,ringrose_salah_2013}. On the other hand, there is a risk of injectivity loss if the fault acts as a barrier for cross-fault fluid flow, causing unwanted pressure build-up within the storage reservoir~\cite{eiken2011,grude2014}. Therefore, understanding the mechanisms of fault flow is an essential aspect of reducing the risk of \co storage in faulted zones.


Fluid flow and pressure communication across and along faults is linked to the underlying fault structure, the host rocks and associated deformation features that are observed in outcrop analogues and laboratory studies. Faults are complex geologic features that includes the fault core and the surrounding damage zone \cite{Bense2013}. 
The fault core is often highly concentrated in space, and is characterized by a multitude of slip surfaces, which can separate several different rock types depending on the faulting mechanism and surrounding host rocks~\cite{braathen2009}. 
The damage zone, which can extend 10s of meters from the fault core, is where the host rock is deformed by fault formation.
For highly porous rocks, such as sedimentary rocks highly relevant for \co storage, an important damage
mechanism is the localization of strain into thin bands that are referred to as deformation bands~\cite{fossen2007}.
Due to grain compaction and crushing during deformation, the permeability of deformation bands in high-porosity formations is usually significantly lower than the surrounding host rock~\cite{torabi2008,sternlof2006}.

Simulation studies are useful tools for quantifying fluid fluid flow. When details of the fault structure are explicitly modeled, complex fluid flow patterns or observed that significantly affect leakage estimates~\cite{zulqarnain2020}. The connection between fault leakage rates and well orientation or injection rate show the importance of field-scale simulation for reliable estimation of fault flow~\cite{newell2020}. The ability to predict \co migration along faults requires extensive fault characterization, the ability to model stress-sensitive parameters, and upscaled flow simulators to provide reliable and efficient results~\cite{snippe2021}.

The damage zone of faults in sedimentary rocks is a key aspect of the overall fault system that affects fluid flow behavior.
Here, the presence of deformation bands can alter fluid flow paths surrounding the fault core. 
Flow can be focused towards the fault if the bands are higher permeability, or hindered if the bands are lower permeability. 
Compared with the fault core, the impact of deformation bands on fluid flow is much less studied. For petroleum reservoirs, the impact of deformation bands on cross-fault fluid flow shows the impact can be significant, e.g., \cite{sternlof2006,Zuluaga2016,Qu2016}. Deformation bands have been attributed to pressure compartmentalization and delayed water breakthrough in hydrocarbon producing reservoirs \cite{Rotevatn2009,Rotevatn2011}.  
For \co storage, studies have shown that deformation bands contribute to capillary trapping of \co which increases storage potential through greater reservoir sweep \cite{torabi2013}.
The scope of studies of deformation bands have thus far been limited to smaller-scale trapping studies~\cite{botter2017impact} or cross-fault fluid flow~\cite{shipton2005} and little is known about how the damage zone might impact the extent fluid leakage vertically along the fault. The impact of deformation bands on pressure build-up near faults in \co storage application is also little understood.

Simulation studies of fault fluid flow have thus far focused on incorporating in the simulation model detailed representations of heterogeneities, including deformation bands in the damage zone \cite{Pourmalek2021,Qu2016}.
While these approaches are useful for creating a detailed representation of fluid flow and gaining important insight, there are severe computational limits for the size of the domain for such a modeling approach. A field-scale simulation model will often have cell sizes in the range of 100 meters, which is significantly larger than the scale of fine-scale fault descriptions, including details in individual deformation bands and their crossings, which is measured at most on the centimeter scale.
Therefore it is necessary to formulate an upscaled geological model for deformation bands, and in a wider context for fault properties in general, that can be used to increase field-scale understanding of the damage zone impact on fault fluid flow.
This can be done using nested models in a multiscale modeling framework \cite{Qu2016}, however, this may result in complex workflows as can be inferred from \cite{Qu2017}.

This paper addresses two remaining challenges with respect to fluid flow in the damage zone of faults in the context of \co storage: 
First, we address the lack of a computationally efficient model that estimates the effective permeability of rocks that contain deformation bands, accounting for both the permeability contrast between deformation bands and the host rock, and the fine-scale geometry of the deformation bands.
Building on initial work, reported in \cite{berge2021}, we propose a new analytical upscaling approach which captures the fine-scale impact of deformation bands on fluid flow in the near-fault region that can be derived from large-scale fault properties. 
Our results show that, first, the analytical upscaling is accurate for complex band geometries, alleviating the need for costly fine-scale computations, and, second, that simpler analytical approaches based on one-dimensional harmonic averages of permeability can substantially overestimate the flow resistance in sets of deformation bands.
The second aspect of our study is the incorporation of the effect of deformation bands on field-scale simulations which model potential leakage of fluids through faults.
We present an approach for incorporating upscaled damage zone permeability into coarse simulation grids in a way that is both reliable and compatible with standard reservoir simulators.
Our methodology is applied to a major fault in a prospective North Sea storage formation. 
We note that the scope of this study is constrained to intrinsic permeability modeling and simulation which is important for understanding pressure development in the vicinity of the fault. We discuss later how this approach can be extended to multiphase flow functions (i.e. relative permeability and capillary pressure). 

The rest of the paper is structured as follows.
The geometry of clusters of deformation bands is introduced in Section \ref{sec:deformation_band_geometry}.
Analytical methods for permeability upscaling in domains which contain deformation bands are derived in Section \ref{sec:perm_upscaling} and the methods are tested against fine-scale numerical simulations in Section \ref{sec:validation}. 
Section \ref{sec:deformation_bandsfield_scale} presents approaches to including the effect of deformation bands in field-scale simulation models, while Section \ref{sec:field_scale_sim} provides simulation results for a prospective storage formation in the North Sea.
A discussion of our findings is provided in Section \ref{sec:discussion}, before concluding remarks are given in Section \ref{sec:conclusion}.

\section{Geometry of deformation bands}\label{sec:deformation_band_geometry}

The permeability of deformation bands can be several orders of magnitude lower than that of the host formation, thus the bands may have a substantial impact on fluid flow from the main formation to the fault core.
Since deformation bands cannot usually be observed directly in a subsurface reservoir they must be treated as stochastic features with properties inferred from a combination of seismic studies and outcrop analogues.
To simplify the modeling, and reduce the computational burden of simulations, we assume that the vertical dimension of the main formation can be disregarded and consider only two-dimensional domains.
Further, the deformation bands are considered straight-line segments, which can be stochastically generated using a marked-point process following \cite{xu2010}, with bands represented by their center locations, orientation and length.
The center locations of the deformation bands are specified via the band density $\rho$ that gives the average number of band centers per unit area. 
The orientation is specified by the rotation $\theta$ from the direction parallel to the fault.
From outcrop studies the distribution of $\theta$ has been found independent of the distance from the fault \cite{berg2005}, and the deformation bands tend to form parallel to the fault; lacking further data we take $\theta$ to be normally distributed with mean 0.
The length of the deformation bands, $l$, is not considered a stochastic variable in this work, in the numerical simulations we will instead sample several values of $l$. In the numerical simulations below, we will vary the band length and the standard deviation of the band rotation to test the sensitivity of the effective permeability on these parameters.

\begin{figure}
    \centering
    \includegraphics[width=0.5\textwidth]{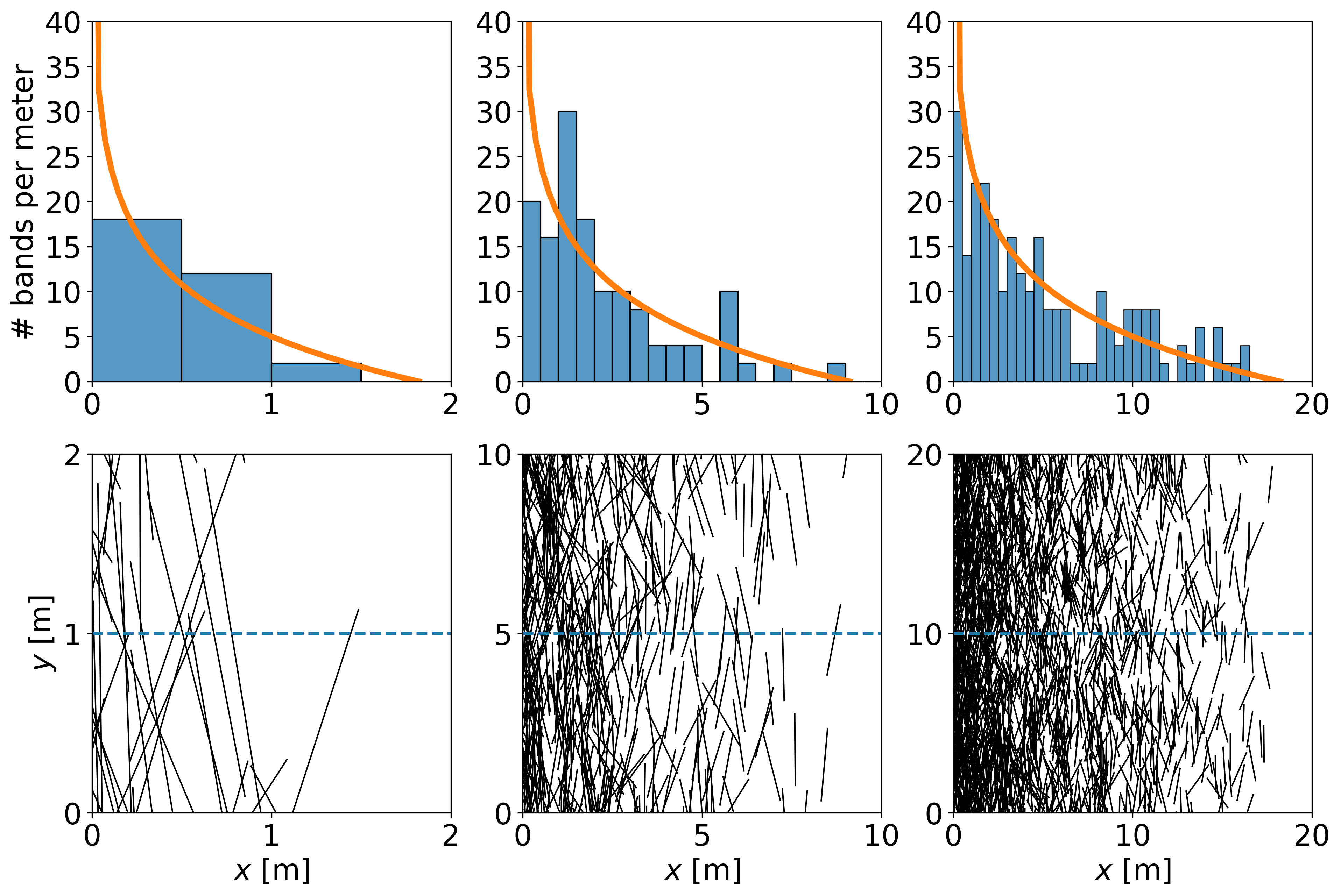}
    \caption{Stochastic realizations of deformation bands in the damage zone of faults with $W_5= 1$ m, $W_5 = 5$ m and $W_5 = 10$ m (from left to right). The density along a scanline follows the logarithmic function in Equation~\eqref{eq:fault_density} and is represented by the orange line. The histograms show the number of deformation bands that intersect the dashed blue line for the three cases.}
    \label{fig:fault_def_band}
\end{figure}

The deformation band density decreases with the distance from the fault. 
Schueller et.al.~\cite{schueller2013} gives a comprehensive study of deformation bands around faults where they study the density profiles of deformation bands. By comparing 106 outcrop scanlines perpendicular to the fault strike they show that there is a logarithmic dependence between the band density and distance from the fault:
\begin{equation}\label{eq:fault_density}
    \rho_x(x) = A + B\ln\left(\frac{x}{1\ \text{m}}\right),
\end{equation}
where the constants $A$ and $B$ are two constants are used to fit the density function to specific faults; we will specify $A$ and $B$ below.
The density $\rho_x$ is the density of deformation bands along a scanline in the $x$-direction (normal to the fault), see Figure~\ref{fig:fault_def_band}. Note that the density of band centers, $\rho$, and the density along a scanline, $\rho_x$, are two different quantities, but they can be derived from each other if the band length and rotation are known. Let $W_5$ be the distance from the fault to where the band density along a scanline perpendicular to the fault is 5 bands per meter, that is $\rho_x(W_5) = 5$ m$^{-1}$. We refer to this length as the damage zone width. Further, we define the average density over the damage zone width as
\begin{equation*}
    \bar{\rho}_{x} = \frac{1}{W_5}\int_0^{W_5} \rho_x(\eta) \, \mathrm{d} \eta.
\end{equation*}
While the average density of the deformation bands over the whole damage zone may vary greatly between faults, the data in~\cite[Fig. 8]{schueller2013} suggests that the average density over the damage zone is independent of the fault throw.
Thus, in this paper we use the median value of the 106 scanlines presented in~\cite{schueller2013}. By setting $\rho_x(W_5)=5$ m$^{-1}$ and $\bar{\rho}_x = 13.33$ m$^{-1}$ we can calculate the constants $A$ and $B$ in Equation~\eqref{eq:fault_density} as
\begin{align*}
    B &= -8.33\ \text{m}^{-1},\\
    A &= \left(5 + 8.33\ln\left(\frac{W_5}{1\ \text{m}}\right)\right)\ \text{m}^{-1}.
\end{align*}

In outcrops, the damage zone width can be found by counting the number of deformation bands per meter, but in a subsurface reservoir, this is usually infeasible, although estimates may be obtained if core samples from the damage zone are available. Thus, the data is to a large degree limited to seismics. 
One of the features that is visible on a seismic resolution is the fault throw, $T$, which can be related to the damage zone width as (see~\cite{schueller2013})
\begin{equation}
    W_5 = 1.74\left(\frac{T}{1\ \text{m}}\right)^{0.43}\ \text{m}.
\label{eq:widthFromThrow}    
\end{equation}

The logarithmic density function, constant average band density, and exponential relation between damage zone width and fault throw allows us to define the band density based on only the fault throw. Together with the assumptions on the band length and rotation we can generate stochastic realizations of the deformation bands in the damage zone of faults. As an example, Figure~\ref{fig:fault_def_band} depicts generated deformation band networks of faults with throw 0.28 m, 12 m and 58 m.

\section{Permeability upscaling}
\label{sec:perm_upscaling}
\begin{figure}
    \centering
    \tiny
    \begin{subfigure}[b]{0.5\textwidth}
    \def\svgwidth{\textwidth}
    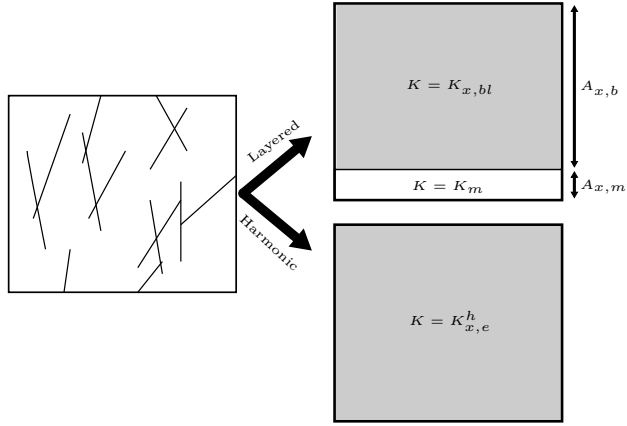
    \end{subfigure}
    \caption{Illustration of two different conceptual representations of a deformation band network. The left figure is the original deformation band network, the top right represents the conceptual layered model and the bottom right represents the conceptual harmonic model, for horizontal fluid flow.}
    \label{fig:conceptualModel}
\end{figure}

It is infeasible to run fine-scale numerical simulations that resolve the geometry of individual bands on a reservoir scale. Instead, we seek to upscale the effect of the deformation bands to an effective permeability of the combined effect of the deformation band permeability and the rock matrix permeability. We consider two different conceptual models, as depicted in Figure~\ref{fig:conceptualModel}. 
A common method to calculate effective permeability is to take the harmonic average of the permeabilities in the deformation bands permeability and the the host rock. 
This method corresponds to a conceptual model where the deformation bands are infinitely long and perpendicular to the flow direction. This conceptual model will be a lower bound on the effective permeability, and the true permeability can be higher because the fluid may for certain geometries flow around the deformation bands. 
An alternative is to upscale the permeability to an equivalent porous medium with two layers of different permeabilities which conceptually represent flow across and around the deformation bands, respectively. 
Both models are considered in this section.

We model the impact of deformation bands on fluid flow by considering an incompressible single-phase fluid. The governing equations are given by Darcy's equation and conservation of mass:
\begin{equation}\label{eq:governing_equations}
    \vec q = -\frac{K}{\mu}\nabla p, \qquad \nabla\cdot \vec q = 0,
\end{equation}
\noindent
where $\vec q$ is the fluid flux, $K$ the permeability, $\mu$ the viscosity and $p$ the pressure. The permeability is assumed isotropic and takes two scalar values, $K_m$ in the host rock and $K_b$ in the deformation bands. In the remainder of the paper the viscosity is set to unit size to simplify notation. In addition, appropriate boundary conditions is assumed to be defined.

\subsection{Effective permeability estimate by harmonic average}
A common method to estimate the effective permeability is to apply the harmonic average. The effective permeability $K_{\alpha,e}^h$ in the direction $\alpha \in \{x, y\}$ is then calculated as:
\begin{equation}\label{eq:KeLowerBound}
\frac{K_{\alpha,e}^h}{K_m} = \frac{1}{(1-\rho_\alpha a)+\frac{\rho_\alpha aK_m}{K_b}}\approx \frac{1}{1+\frac{\rho_\alpha aK_m}{K_b} }.
\end{equation}
Here, $\rho_\alpha$ represents the band density along a scanline in the direction $\alpha$ and $a$ is the aperture of the deformation bands. In this conceptual model it is assumed that the fluid crosses all deformation bands along the scanline. Thus, this will represent a lower bound on the effective permeability that can can be obtained, e.g., if the deformation bands are infinitely long.

The approximation of the denominator in the last step in Equation~\eqref{eq:KeLowerBound} makes the permeability reduction only dependent on the dimensionless quantity $\rho_\alpha aK_m/K_b$.
The error of this approximation is small if $\rho_\alpha a\ll 1$. 
This will often be the case, for instance, of the 106 outcrop scanlines presented in~\cite{schueller2013}, the maximum band density of all scanlines was less than 110 bands per meter. 
The mean maximum value per scanlines is 34 bands per meter.
Thus, the approximation error is small if the bands have an aperture of, say, $a\approx 1$ mm.

\subsection{Effective permeability estimate by two-layer model}\label{sec:harmonic_avg}
In this sub-section we develop an approach for estimation of the effective permeability which allows for partial diversion of the flow around deformation bands.
The conceptual model underlying the upscaling, depicted in Figure~\ref{fig:conceptualModel} for fluid flow in the $x$-direction, represents
 flow across and around bands with two separate layers with different permeabilities.
In the derivation of the upscaled permeability we make the simplifying assumption that the band density, $\rho$, is constant in the domain.
While this contradicts the density relation given by Equation~\eqref{eq:fault_density}, it allows for the derivation of analytical expressions for the upscaled permeability.
We show in Section~\ref{sec:res_fault} that the approach can be extended to the vicinity of faults where the density is dependent on the distance from the fault.

We assign the host rock matrix permeability, $K_m$, as the permeability in the host rock layer. The band layer consists of both deformation bands and host rock. The permeability of this layer, $K_{\alpha,bl}$, in the direction $\alpha$, is calculated as the harmonic average of the deformation bands and the host rock, and applying the approximation of small aperture:
\begin{equation}\label{eq:effective_perm_band_layer}
\frac{K_{\alpha,bl}}{K_m} = \frac{1}{1+ c_\alpha\frac{a\rho_\alpha K_m}{K_b} }.
\end{equation}
Here, the constant $c_\alpha$ is a tuning parameter that can be interpreted as the ratio between the number of bands in the layer the fluid must cross and the total number of bands within the band layer. The effective permeability of a two-layer model can then be calculated as
\begin{equation}\label{eq:effective_perm}
    K_{\alpha, e}^l = \frac{K_{\alpha,bl}A_{\alpha, b} + K_mA_{\alpha,m}}{A_{\alpha, b}+A_{\alpha,m}},
\end{equation}
where $A_{\alpha,b}$ and $A_{\alpha,m}$ are the cross sectional areas of layers associated with fluid crossing the bands and fluid flowing in the host rock, respectively (see Figure~\ref{fig:conceptualModel}). We note that the harmonic mean in Equation~\eqref{eq:KeLowerBound} corresponds to $A_{\alpha,m}=0$ and $c_{\alpha}=1$.

\begin{figure}
    \centering    
    \begin{subfigure}[b]{0.5\textwidth}
    \def\svgwidth{\textwidth}
    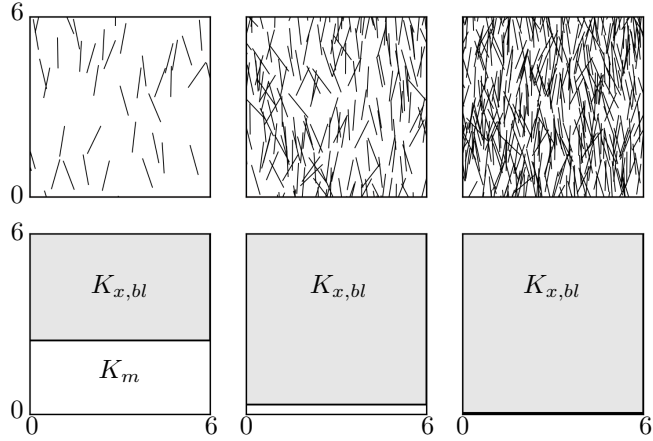
    \end{subfigure}
    \caption{The figure shows deformation band networks with $\rho_x = 1,\ 5,\ 10$ m$^{-1}$ (left to right), and the corresponding conceptual layered models are shown in the bottom row. The length of the deformation bands is $l=1$ m and the rotation, $\theta$, is normally distributed with standard deviation $\sigma=\pi/12$.}
    \label{fig:network_and_A}
\end{figure}

The task now is to calculate approximations of the cross sectional areas of the band layer, $A_{\alpha,m}$, and the host rock layer, $A_{\alpha,b}$.
When the band density is small, the domain mainly consists of individual bands that do not intersect. In this case we can estimate the areas, $A_{\alpha,b}$ and $A_{\alpha, m}$ as the band length and the distance between the bands. However, when the band density increases the domain will be dominated by deformation bands that intersect and forming clusters, see Figure~\ref{fig:network_and_A}. In that case, we estimate the area available for flow across the deformation bands as the cluster size, while the area available for flow around the deformation bands is still the area between deformation bands. 

The below calculation of $A_{\alpha,m}$ and $A_{\alpha,b}$ proceeds in three steps: We first derive the probability of intersection between two arbitrary bands, and use this to estimate the expected number of intersections for a given band. This can finally be used to estimate the areas of interest.

\subsubsection{Probability of intersection between two bands}
Consider two arbitrary deformation bands. Let the rotation of the deformation bands be normally distributed with mean $0$ and standard deviation $\sigma$.
Denote by $\theta_1 \sim \mathcal N(0, \sigma^2)$ and $\theta_2 \sim \mathcal N(0, \sigma^2)$ the rotation of the first and second deformation band.  Without loss of generality (when we later let the domain size go to infinity), we define a new coordinate system such that the center of the first band is located at origin and such that the $y$-axis is parallel to the first band. 
The rotations of the two bands in the new coordinate system are defined by $\theta_1' = 0$ and
$\theta_2'=\theta_2 - \theta_1$.  We assume that the position of the second band center is independent and uniformly distributed in the domain $[-L_x/2, L_x2]\times[-L_y/2, L_y/2]$. The center position of the second  deformation band can then be described by the $x$- and $y$-coordinates that are uniformly distributed:
\begin{equation*}
    x\sim U\left(-\frac{L_x}{2}, \frac{L_x}{2}\right),\quad y\sim U\left(-\frac{L_y}{2}, \frac{L_y}{2}\right).
\end{equation*} 
Assuming constant length of the two bands, $l$, the two line segments representing the bands can be parameterized in the new
coordinate system as
\begin{align*}
  x_1 &= 0, &&y_1 = t_1 \frac{l}{2}, &&& t_1 \in [-1, 1],\\
  x_2 &= x + t_2\frac{l}{2}\sin(\theta_2'), &&y_2 = y + t_2\frac{l}{2}\cos(\theta_2'), &&& t_2 \in [-1, 1].
\end{align*}
The lines defined by the parameterization intersect at the point where $x_1 = x_2$, and $y_1 =
y_2$. Using these two equations and solving for $t_1$ and $t_2$ we obtain
\begin{equation}\label{eq:intersection_param}
  t_1 = \frac{2y}{l} - \frac{2x\cos(\theta_2')}{l\sin(\theta_2')}, \quad t_2 = -\frac{2x}{l\sin(\theta_2')}.
\end{equation}
The two line segments intersect iff $t_1\in[-1, 1]$ and $t_2\in[-1,1]$.
Assume $L_x>l$, $L_y > l$, that is, the domain is sufficiently large to contain the band length and thus all possible intersections. 
From Equation~\eqref{eq:intersection_param} we obtain that $|t_1|\le 1$ and $|t_2|\le 1$ for 
\begin{align*}
    -\frac{l}{2}|\sin(\theta_2')| \le x\le \frac{l}{2} |\sin(\theta_2')|,\qquad
    \left|\frac{\cos(\theta_2')}{\sin(\theta_2')}\right|x - \frac{1}{2}l \le y \le \left|\frac{\cos(\theta_2')}{\sin(\theta_2')}\right|x + \frac{1}{2}l.
\end{align*}
Thus, the region where the deformation bands intersect forms a parallelogram with area $l^2|\sin(\theta_2')|$. If we are given the two rotations $\theta_1$ and $\theta_2$, the probability of the two deformation bands intersecting is:
\begin{equation*}
    P(\text{intersect}| \theta_1,\theta_2) = \frac{l^2|\sin(\theta_2 - \theta_1)|}{L_xL_y}.
\end{equation*}

The probability of the two arbitrary bands intersecting can then be calculated by applying the law of total probability:
\begin{equation}\label{eq:prob_two_bands_intersect}
  P(\text{intersect}) = \frac{l^2}{L_xL_y}\int_{-\infty}^\infty \int_{-\infty}^\infty 
  \frac{|\sin(\theta-\xi)|}{2\pi\sigma^2}\exp\left(-\frac{\theta^2}{2\sigma^2}\right)\exp\left(-\frac{\xi^2}{2\sigma^2}\right)\, \mathrm{d} \theta\, \mathrm{d} \xi,
\end{equation}
which can approximated numerically.

\subsubsection{Expected number of intersection for a single band} \label{sec:number_of_intersections}
The number of intersections for a given band can be derived from the probability of intersection between two  bands.
Let $m + 1$ be the number of deformation bands in the domain and let $I$ be the number of intersections the first band have with the remaining bands. The probability that none of these
intersecting the first band is
\begin{equation*}
  P(I = 0) = (1 - P(\text{intersect}))^m.
\end{equation*}
We define the band density as $\rho = m / (L_xL_y)$. By fixing the density $\rho$ and
letting the domain size go to infinity the probability of the first band having zero
intersections is
\begin{equation*}
  P(I=0) = \left(1 - \frac{\rho L_xL_yP(\text{intersect})}{\rho L_xL_y}\right)^{L_xL_y\rho}\overset{L_x=L_y\rightarrow\inf}{=}
  \exp\left(-\rho L_xL_yP(\text{intersect})\right).
\end{equation*}
Note that the product $L_xL_yP(\text{intersect})$ stays bounded as the domain grows to
infinity because the area $L_xL_y$ cancels with the factor $1/(L_xL_y)$ in $P(\text{intersect})$. 
Inserting Equation~\eqref{eq:prob_two_bands_intersect} into the expression above we obtain:
\begin{equation}\label{eq:prob_zero_int}
P(I=0)=\exp \left(-l^2\rho\int_{-\infty}^\infty \int_{-\infty}^\infty 
  \frac{|\sin(\theta-\xi)|}{2\pi\sigma^2}\exp\left(-\frac{\theta^2}{2\sigma^2}\right)\exp\left(-\frac{\xi^2}{2\sigma^2}\right)\, \mathrm{d} \theta\, \mathrm{d} \xi\right),
\end{equation}
which we approximate numerically.

When the position and rotation of all deformation bands are independent, the number of intersections for each deformation band is
Poisson distributed:
\begin{equation*}
  f(k;\nu) = P(I = k) = \frac{\nu^k\exp(-\nu)}{k!}.
\end{equation*}
Equation~\eqref{eq:prob_zero_int} gives the probability of zero intersections, i.e., $f(0;\nu)$, thus, we can obtain the expected value for the number of intersections as
\begin{equation}\label{eq:expected_number_of_cross}
  \nu=E(I) = -\log(P(I = 0)).
\end{equation}
\begin{figure}
    \centering
    \includegraphics[width=0.5\textwidth,trim={0 0 0 1.5cm},clip]{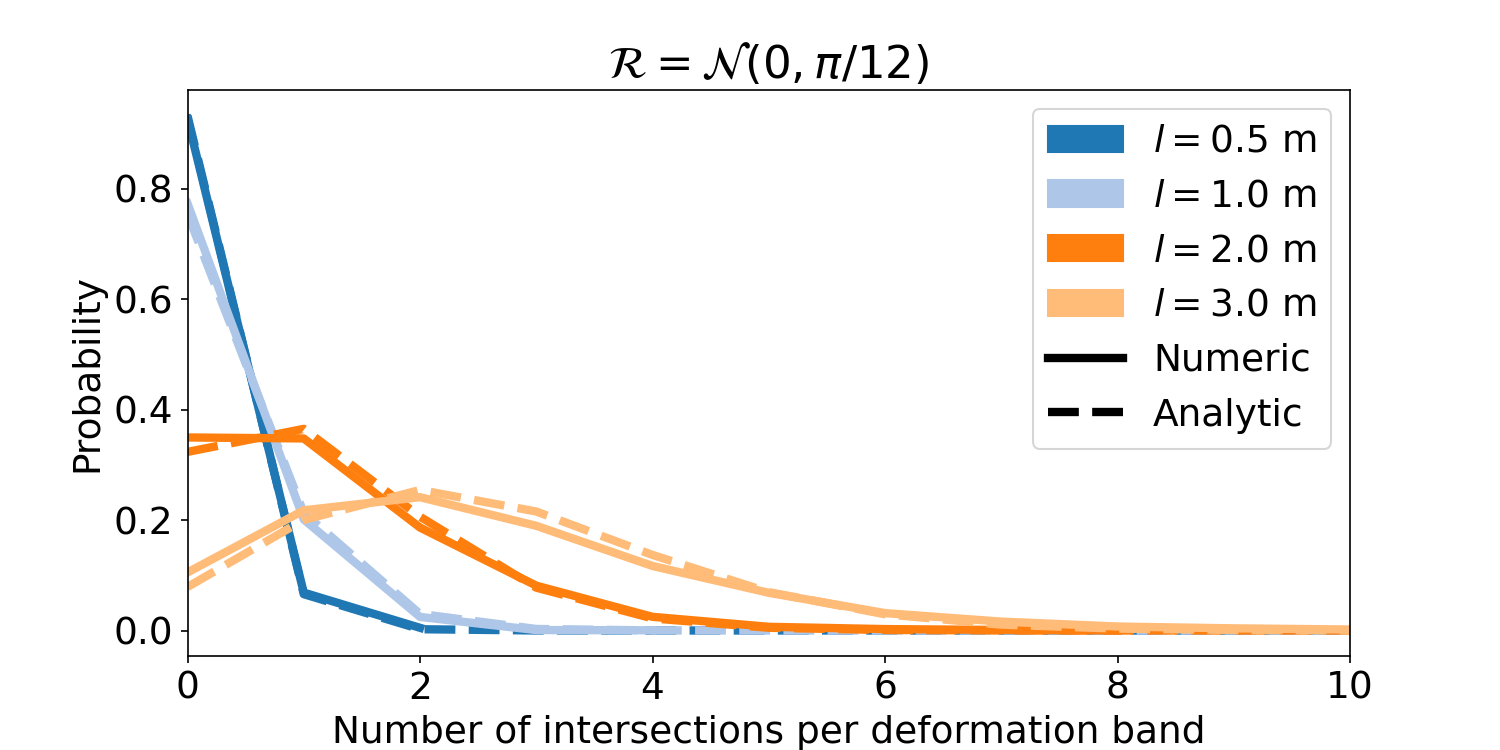}
    \caption{The probability distribution of the number of intersections, $I$, a deformation band has. Each color corresponds to a different band length. Numerical estimations of the probability are represented by the solid lines, while the dashed lines show the probabilities of the Poisson distribution with an expected value given by Equation~\eqref{eq:expected_number_of_cross}. The other parameters used to generate the deformation band networks are $\{\sigma, \rho\} = \{\pi/12, 1 \text{ m}^{-2}\}$.}
    \label{fig:number_of_intersections}
\end{figure}
Figure~\ref{fig:number_of_intersections} compares the Poisson distribution with an expected
value calculated from the analytical expression given in
Equation~\eqref{eq:expected_number_of_cross} to a numerical calculation of the
distributions. The numerical distributions are obtained by placing 10 000 bands randomly in a domain, and counting the the average number of intersections per
band.

\subsubsection{Areas available for flow}
As seen in Figure~\ref{fig:network_and_A}, the deformation bands may form chains of intersecting deformation bands. In the layered conceptual model this will increase the area of the band layer, $A_{\alpha, b}$. While we have an analytic expression for the expected number of intersections a single deformation band has, calculating the probability of the number of bands in a chain is more difficult. All chains have at least one deformation band. We estimate the probability of a chain having additional $n$ deformation bands as $P(I\ge 1)^n$, that is, the probability of randomly drawing $n$ bands in a row with at least 1 intersection. Thus, the expected value of the number of additional deformation bands in a chain is approximated by:
\begin{equation*}
    E(C) = \sum_{n=1}^\infty nP(I\ge1)^n = \frac{P(I\ge 1)}{(1 - P(I\ge 1))^2}.
\end{equation*}
The first band in the chain extends the chain with a length $l_{\alpha}$, in the direction perpendicular to $\alpha$.
Because the position of deformation bands is independent and uniformly distributed, we assume that the length of a chain scales as the square root of the number of additional bands in the chain. Thus, we calculate the area available for flow in the band layer as
\begin{equation*}
    A_{\alpha, b}=l_\alpha^\top+\sqrt{\frac{l_\alpha^\top}{2}  E(C)}.
\end{equation*}
The expected band length in direction perpendicular to $\alpha$ is calculated by assuming $\sigma \ll 1$:
\begin{equation*}
    l_x^\top= l\cos\left(\sigma\sqrt{\frac{2}{\pi}}\right),\quad l_y^\top = l\sin\left(\sigma\sqrt{\frac{2}{\pi}}\right),\quad \rho_\alpha = \rho l_\alpha^\top,
\end{equation*}
here, the value $\sigma \sqrt{2/\pi}$ is the expected absolute value of the rotation, $E(|\theta|)$.

For the host rock layer in the conceptual model, the area $A_{\alpha, m}$ is set inversely proportional to the band density:
\begin{equation*}
    A_{\alpha,m} = \frac{1}{\rho_\alpha}.
\end{equation*}

As an example, Figure~\ref{fig:network_and_A} shows three different deformation band networks and their corresponding conceptual models. The three networks are generated with parameters $l=1$ m, $\sigma =\pi/12$ and $\rho_x = \{1,\ 5,\ 10\}$ m$^{-1}$. For the first case, $\rho_x=1$ m$^{-1}$, the network mainly consists of individual bands, but as the density increases, the deformation bands form long chains of crossing bands that greatly reduces the area available for fluid flow around the deformation bands.

\section{Validation against fine-scale numerical simulations}
\label{sec:validation}
\begin{figure}
    \centering    
    \newcommand{\figWidth}{0.24\textwidth}
    \begin{minipage}[c]{0.96\textwidth}
    \begin{subfigure}[b]{\figWidth}
    \includegraphics[width=\textwidth]{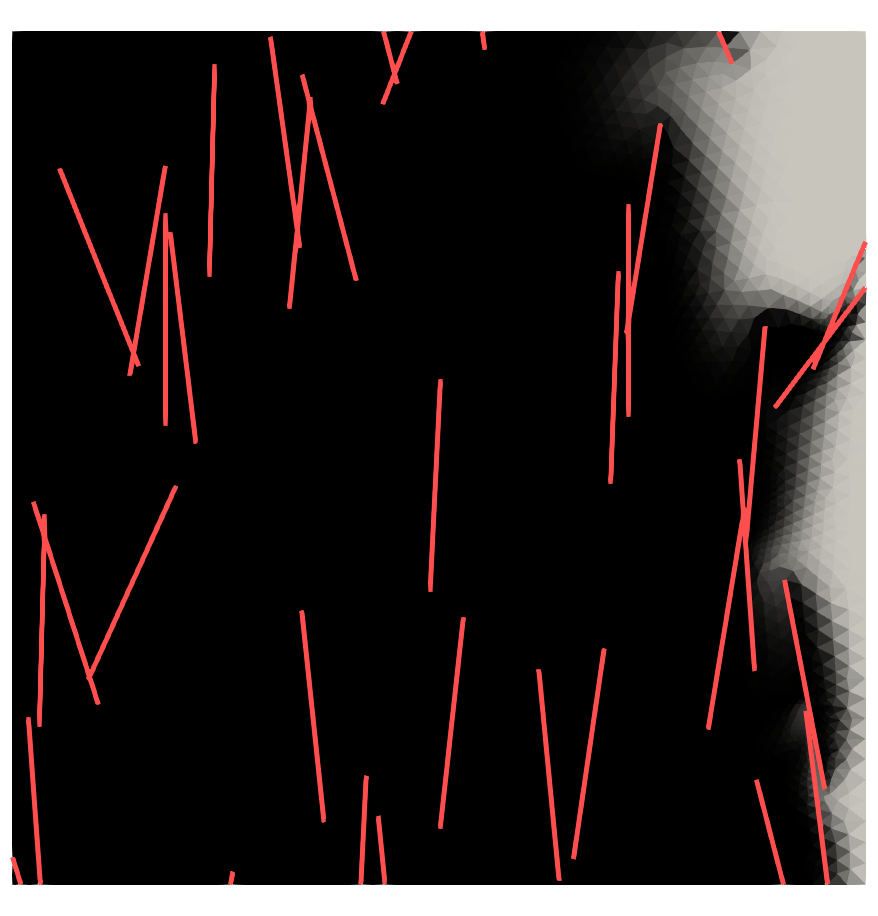}
    \subcaption{$\frac{tK_m \Delta p}{L_x^2\phi \mu} = 0.25$}
    \end{subfigure}
    \begin{subfigure}[b]{\figWidth}
    \includegraphics[width=\textwidth]{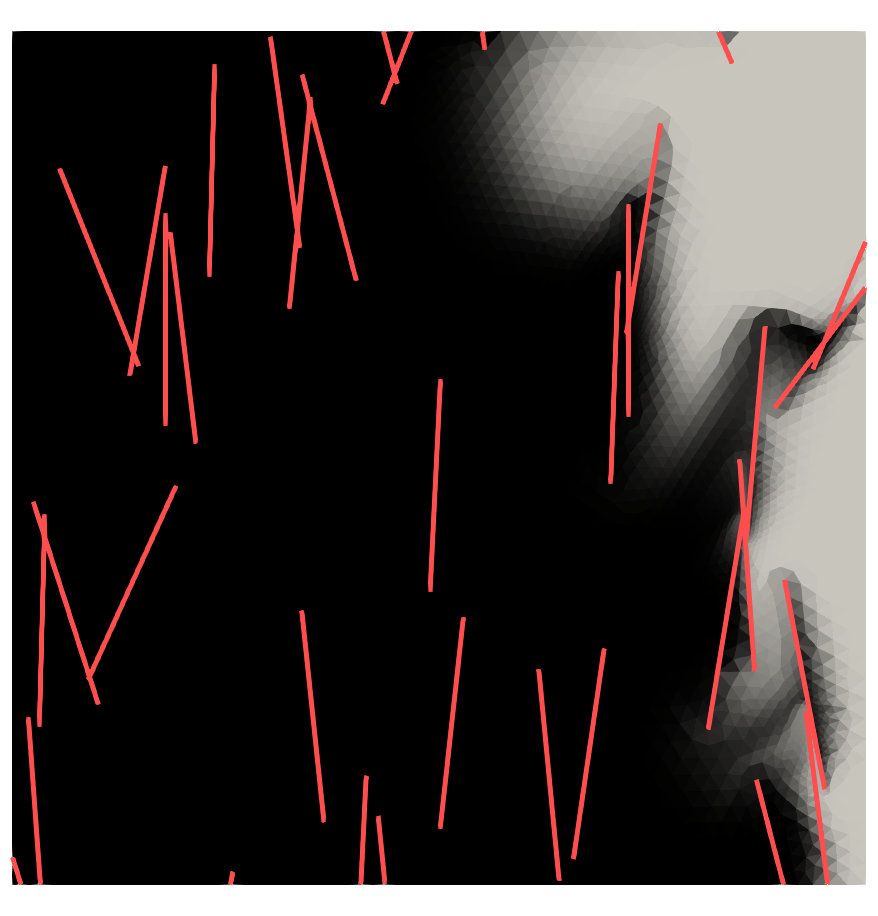}
    \subcaption{$\frac{tK_m \Delta p}{L_x^2\phi \mu} = 0.5$}
    \end{subfigure}
    \begin{subfigure}[b]{\figWidth}
    \includegraphics[width=\textwidth]{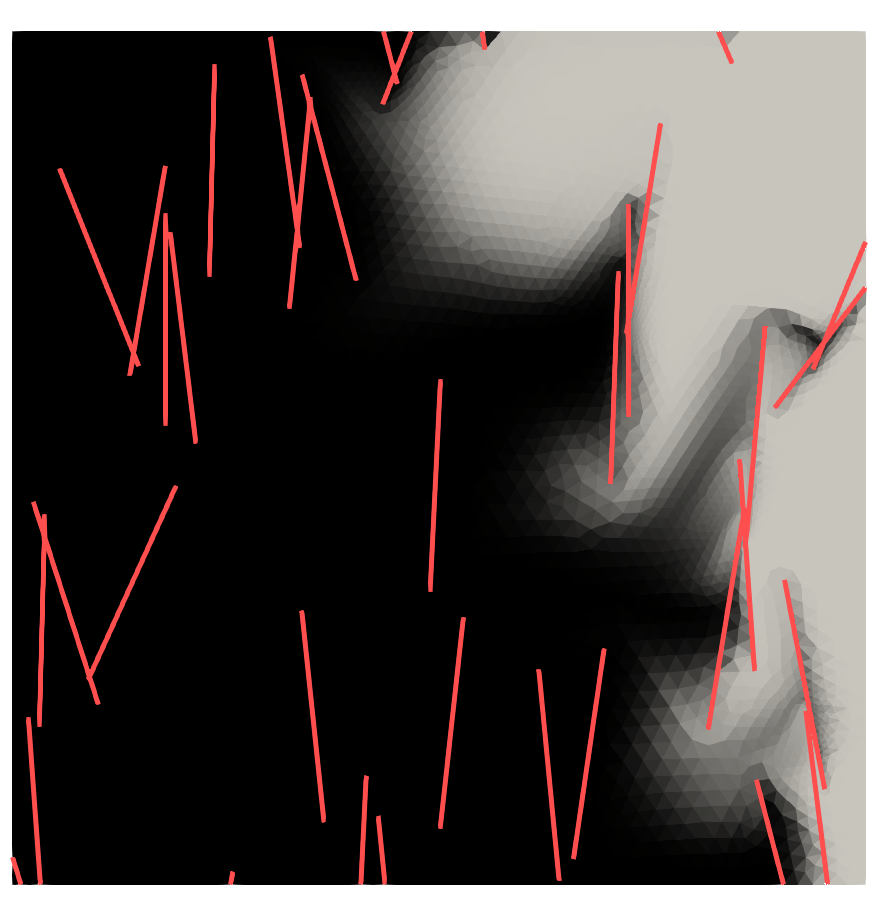}
    \subcaption{$\frac{tK_m \Delta p}{L_x^2\phi \mu} = 0.75$}
    \end{subfigure}
    \begin{subfigure}[b]{\figWidth}
    \includegraphics[width=\textwidth]{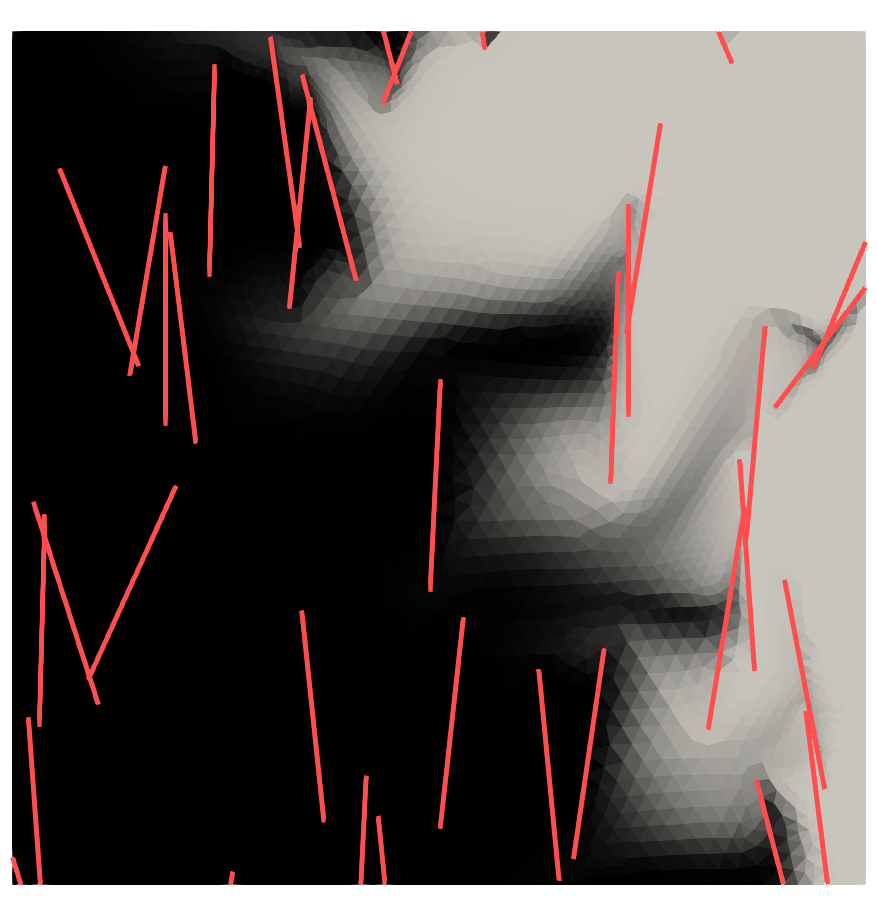}
    \subcaption{$\frac{tK_m \Delta p}{L_x^2\phi \mu} = 1$}
    \end{subfigure}
    \end{minipage}
    \begin{minipage}[c]{0.03\textwidth}
    \def\svgwidth{\textwidth}
\begingroup%
  \makeatletter%
  \providecommand\color[2][]{%
    \errmessage{(Inkscape) Color is used for the text in Inkscape, but the package 'color.sty' is not loaded}%
    \renewcommand\color[2][]{}%
  }%
  \providecommand\transparent[1]{%
    \errmessage{(Inkscape) Transparency is used (non-zero) for the text in Inkscape, but the package 'transparent.sty' is not loaded}%
    \renewcommand\transparent[1]{}%
  }%
  \providecommand\rotatebox[2]{#2}%
  \newcommand*\fsize{\dimexpr\f@size pt\relax}%
  \newcommand*\lineheight[1]{\fontsize{\fsize}{#1\fsize}\selectfont}%
  \ifx\svgwidth\undefined%
    \setlength{\unitlength}{99.91316986bp}%
    \ifx\svgscale\undefined%
      \relax%
    \else%
      \setlength{\unitlength}{\unitlength * \real{\svgscale}}%
    \fi%
  \else%
    \setlength{\unitlength}{\svgwidth}%
  \fi%
  \global\let\svgwidth\undefined%
  \global\let\svgscale\undefined%
  \makeatother%
  \begin{picture}(1,5.87597019)%
    \lineheight{1}%
    \setlength\tabcolsep{0pt}%
    \put(0,0){\includegraphics[width=\unitlength,page=1]{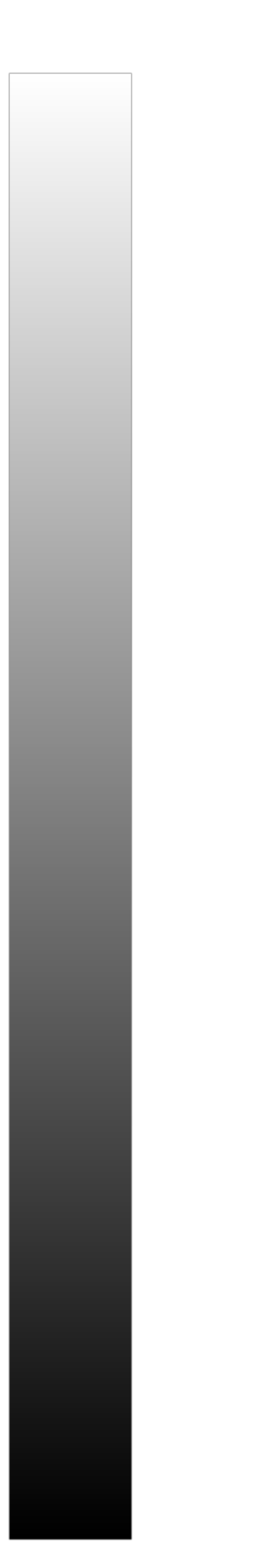}}%
    \put(0.52325594,0.03875955){\color[rgb]{0,0,0}\makebox(0,0)[lt]{\lineheight{1.25}\smash{\begin{tabular}[t]{l}0\end{tabular}}}}%
    \put(0.59108549,5.56976849){\color[rgb]{0,0,0}\makebox(0,0)[lt]{\lineheight{1.25}\smash{\begin{tabular}[t]{l}1\end{tabular}}}}%
    \put(0.63565899,3.13953537){\color[rgb]{0,0,0}\makebox(0,0)[lt]{\lineheight{1.25}\smash{\begin{tabular}[t]{l}$c$\end{tabular}}}}%
    \put(0,0){\includegraphics[width=\unitlength,page=2]{color_bar.pdf}}%
  \end{picture}%
\endgroup%

    \end{minipage}
    \caption{Tracer transport in a domain of size $L_x\times L_x = 4$ m $\times$ 4 m for four snapshots of the dimensionless time, where $\mu$ is the viscosity, $\phi$ the porosity and $\Delta p$ the pressure drop over the domain. The red lines represent the deformation bands. The parameters used in the simulation are $\{\rho_x, l, \sigma, K_b / (aK_m)\} = \{2\  \text{m}^{-1}, 1\ \text{m}, \pi/12, 10^{-1}\ \text{m}^{-1}\}$. Initially, the domain contains no tracer, $c=0$, but the tracer gradually fills the domain, $c=1$.}
    \label{fig:tracer_transport}
\end{figure}

To test the two different upscaled models, we will compare the effective permeabilities calculated by the two models to the effective permeability obtained by fine-scale numerical simulations. In the fine-scale simulations, the deformation bands are included explicitly in the simulation domain which allows us to simulate the intricate flow patterns that can arise due to the deformation bands. An example of the interaction between the deformation bands and the fluid flow is shown in Figure~\ref{fig:tracer_transport}. The figure depicts the simulation of a tracer through a domain of size $4$ m $\times$ 4 m.

Two different setups are considered. In the first, we consider a scanline band density $\rho_x$ that is constant in the domain. In the second, we let $\rho_x$ vary as defined by Equation~\eqref{eq:fault_density}.
The scaled permeability ratio is defined as the band permeability divided by the aperture and the rock matrix permeability. Table~\ref{tab:range_of_parameters} shows typical values for the aperture $a$ and the permeability contrast between the deformation bands and rock matrix. To cover the range of plausible scaled permeability ratios, it is varied four orders of magnitudes for all test cases:
\begin{equation*}
    \frac{K_b}{aK_m} = \{10^{-2},\ 10^{-1},\   10^0,\ 10^1\}\ \text{m}^{-1}.
\end{equation*}
To relate these values to a permeability ratio, we may assume an aperture of $a=1$ mm which gives:
\begin{equation*}
    \frac{K_b}{K_m} = \{10^{-5},\ 10^{-4},\   10^{-3},\ 10^{-2}\}.
\end{equation*}
\begin{table}
    \centering
    \caption{Range of parameters that one typically can observe for deformation bands~\cite{fossen2007}. Using the maximum aperture and largest permeability contrast we obtain the scaled permeability ratio $\frac{K_b}{aK_m}= 10^{-3}$ m$^{-1}$. }
    \label{tab:range_of_parameters}
    \begin{tabular}{c c c} 
     \hline
      & Min & Max \\ [0.5ex] 
     \hline
     $a$ & 0.1 mm & 10 mm \\ 
     $\frac{K_b}{K_m}$ & $10^{-5}$ & 1.0\\ [1ex]
     \hline
    \end{tabular}
\end{table}

The numerical simulations are run on rectangular computational domains. The sizes of the domains are chosen such that a doubling of the sizes do not change the numerical calculation of the effective permeability. The height of the computational domain, $L_y$, is set equal to 8 times the band length. The domain width, $L_x$, differ between the two cases. In the constant density case, the width equals $4$ times the band length, while for the fault-case the width is the distance (from the fault) at which  the band density is zero, that is,
\begin{equation}\label{eq:domain_width}
    \rho_x(L_x)=0 \rightarrow L_x = \exp\left(-\frac{A}{B}\right).
\end{equation}

The rotation of the bands, $\theta\sim\mathcal N(0,\sigma^2 )$, is assumed to be normally distributed with standard deviation $\sigma$ and mean 0. The effective permeability of the layered model given in Equation~\eqref{eq:effective_perm} requires a tuning parameter $c_\alpha$ to be set. In all examples we use
\begin{equation*}
    c_x=1,\qquad c_y=\sin\left(\sigma\sqrt{\frac{2}{\pi}}\right).
\end{equation*}

\subsection{Fine-scale numerical method}
We make two comments on our numerical approach: First, flow in the tangential direction of the deformation bands is neglected in our simulations, since the band aperture $a$ is typically measured in millimeters and their permeability $K_b$ is lower than that of the host rock. Second, the computational grids are constructed such that the faces of the grids conform to the deformation bands. 


The governing equations~\eqref{eq:governing_equations} are discretized using a finite volume scheme and applying the Two-Point Flux Approximation (TPFA) to calculate the flux over each face. The TPFA method is sufficiently accurate for our objectives, as we are only interested in calculating the reduction in effective permeability relative to the homogeneous case. 
The TPFA method calculates the flux over a face $\psi$ based on the pressure difference in the neighbour cells denoted by $L$ and $R$:
\begin{equation*}
    F_\psi = -T_\psi(p_R - p_L).
\end{equation*}
Here, $T_\psi$ is the face transmissibility, and the the cell center pressures of the two cells are denoted by $p_L$ and $p_R$. The standard way to calculate the face transmissibility is:
\begin{equation}
\label{eq:trm}
    T_\psi = \frac{T_{\psi L}T_{\psi R}}{T_{\psi L} + T_{\psi R}},\quad
    T_{\psi k} = \frac{K_mA_\psi \vec n_{\psi k} \cdot \vec d_{\psi k}}{\vec d_{\psi k}\cdot \vec d_{\psi k}}, \quad k\in \{L, R\}.
\end{equation}
The face area is denoted by $A_\psi$, the vector from the cell center to the face center is denoted by $\vec d_{\psi k}$, and the unit normal (pointing out of cell $k$) of the face is denoted by $\vec n_{\psi k}$. The transmissibility in Equation~\eqref{eq:trm} is used for all faces that do not lie on a deformation band. 

The deformation bands can be included in the TPFA method by a simple modification to the transmissibility if we neglect the tangential flow in the deformation bands, assume a small band aperture, and use a conforming grid. This allows us to include the effect of deformation bands without having to introduce numerical cells inside the deformation bands. For the faces on a deformation band, the half transmissibilities of the neighbour cells are calculated by taking into account the reduction of  permeability due to the deformation band on the face:
\begin{equation*}
    T_\psi = \frac{T_{b\psi L}T_{b\psi R}}{T_{b\psi L} + T_{b\psi R}},\quad
    T_{b\psi k} = \frac{K_bT_{\psi k}}{K_b + \frac{T_{\psi k}a}{2A_\psi}},
     \quad k\in \{L, R\}.
\end{equation*}
This face transmissibility is used for all faces that conform to deformation bands in the computational grid.

\subsection{Case with constant band density}
\label{sec:res_homogen}
In this test case, the spatial variations of the band density function is constant in the whole domain. In all simulations in this subsection, the standard deviation of the band rotation is fixed to $\sigma = \pi / 12$. The band length, $l$, the band density along a scanline, $\rho_x$, and the scaled permeability ratio $K_b/(aK_m)$ are varied.

\begin{figure}
    \centering
    \newcommand{\figWidth}{0.24\textwidth}
    \includegraphics[width=0.3\textwidth]{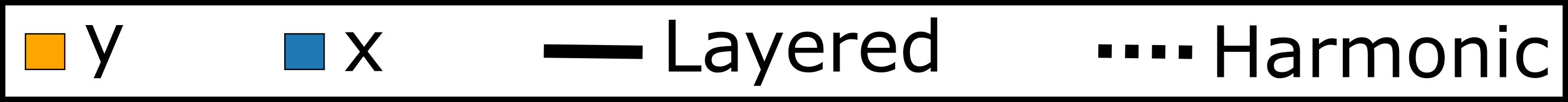}\\
    \begin{subfigure}[b]{\figWidth}
    \includegraphics[width=\textwidth]{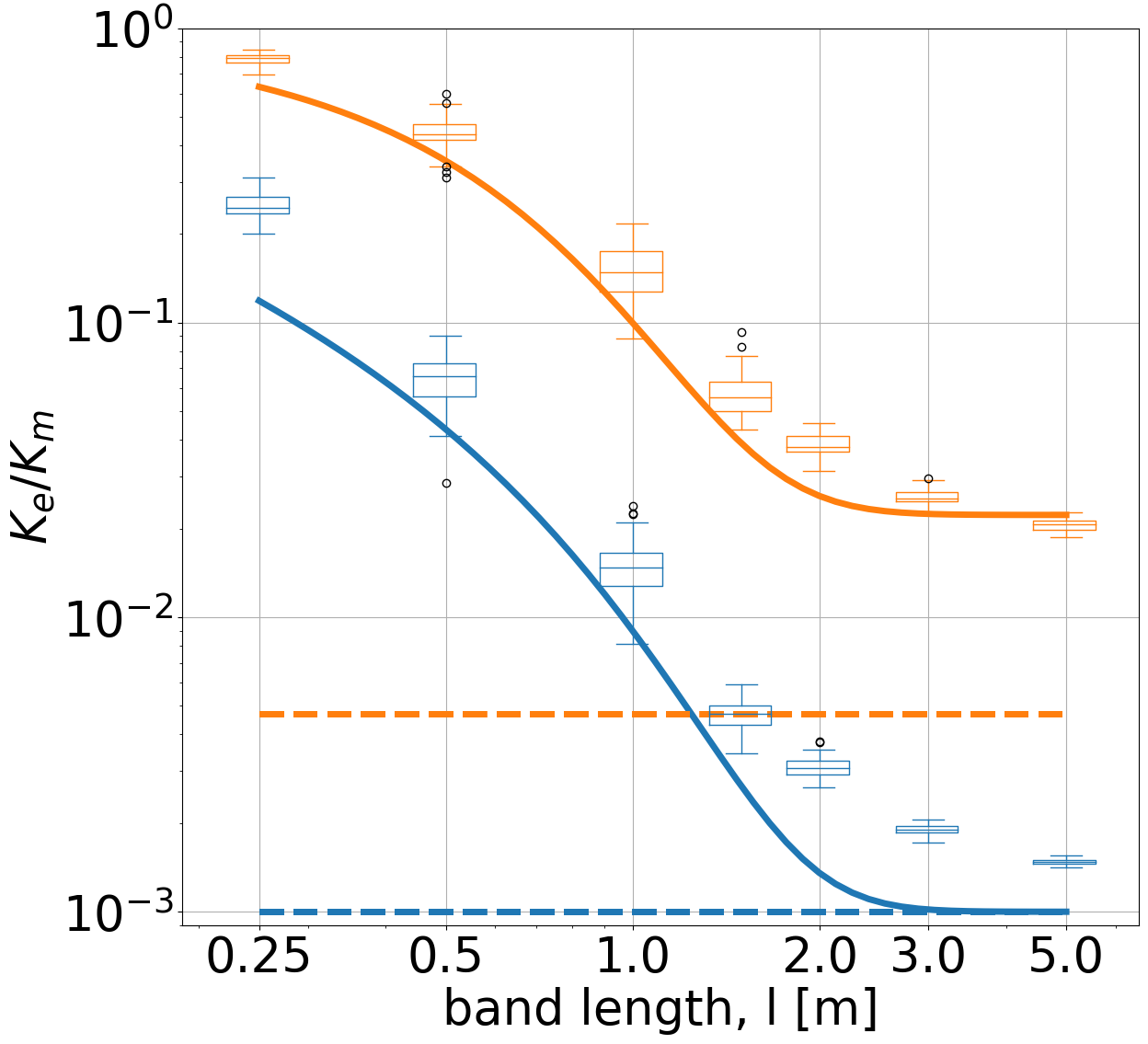}
    \caption{$\frac{K_b}{aK_m} = 10^{-2}$ m$^{-1}$}
    \end{subfigure}
    \begin{subfigure}[b]{\figWidth}
    \includegraphics[width=\textwidth]{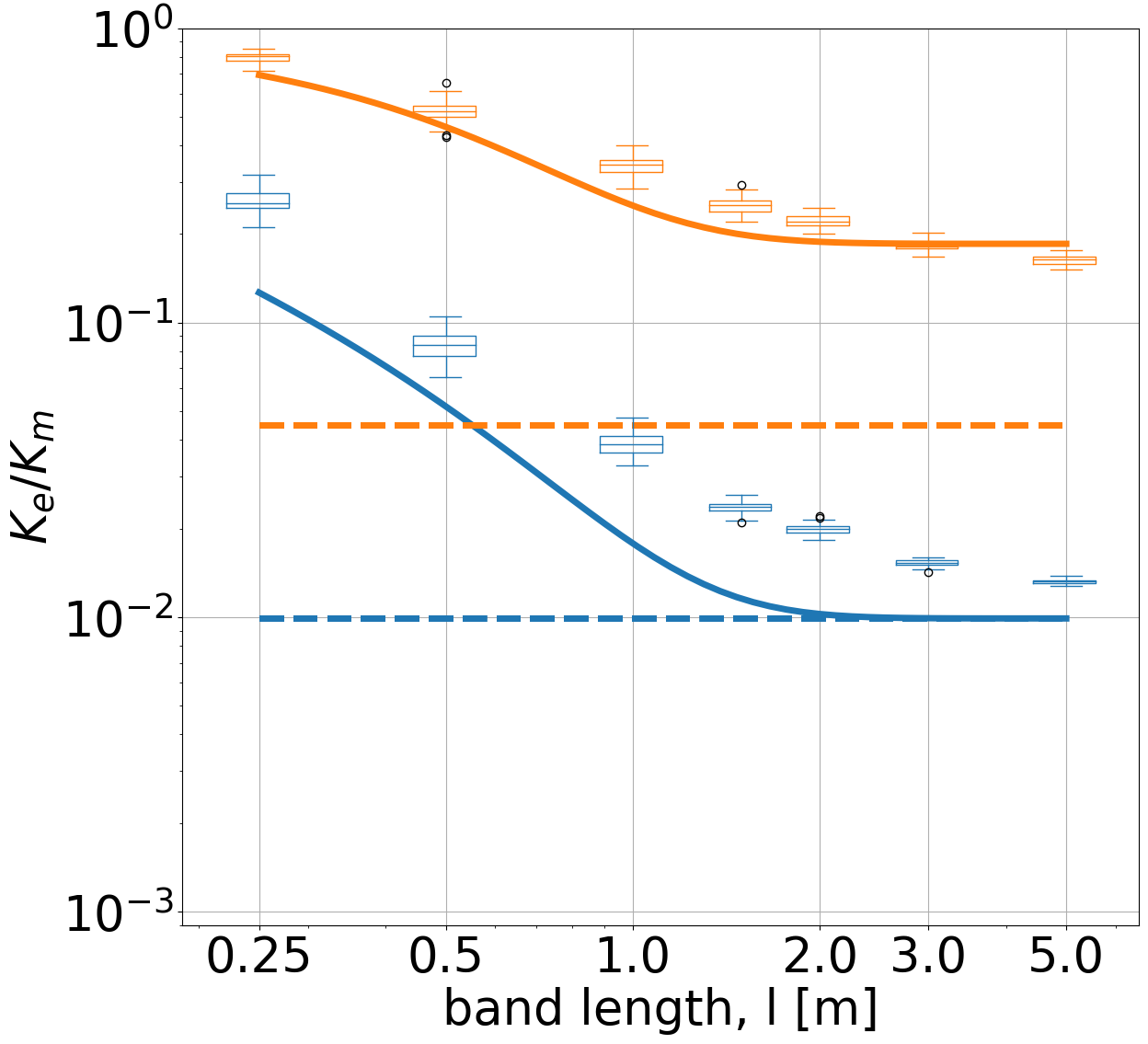}
    \caption{$\frac{K_b}{aK_m} = 10^{-1}$  m$^{-1}$}
    \end{subfigure}
    \begin{subfigure}[b]{\figWidth}
    \includegraphics[width=\textwidth]{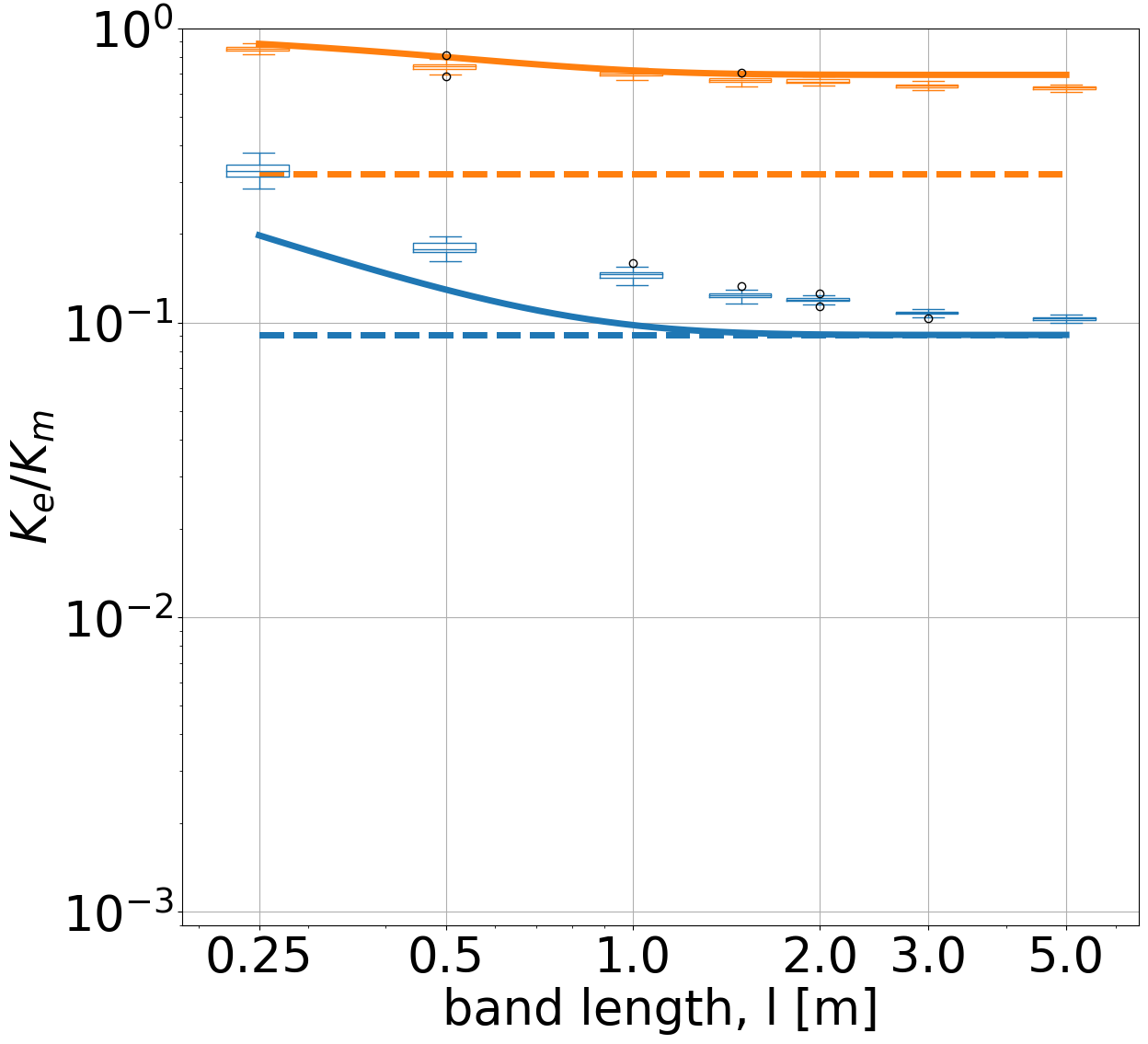}
    \caption{$\frac{K_b}{aK_m} = 10^{0}$ m$^{-1}$}
    \end{subfigure}
    \begin{subfigure}[b]{\figWidth}
    \includegraphics[width=\textwidth]{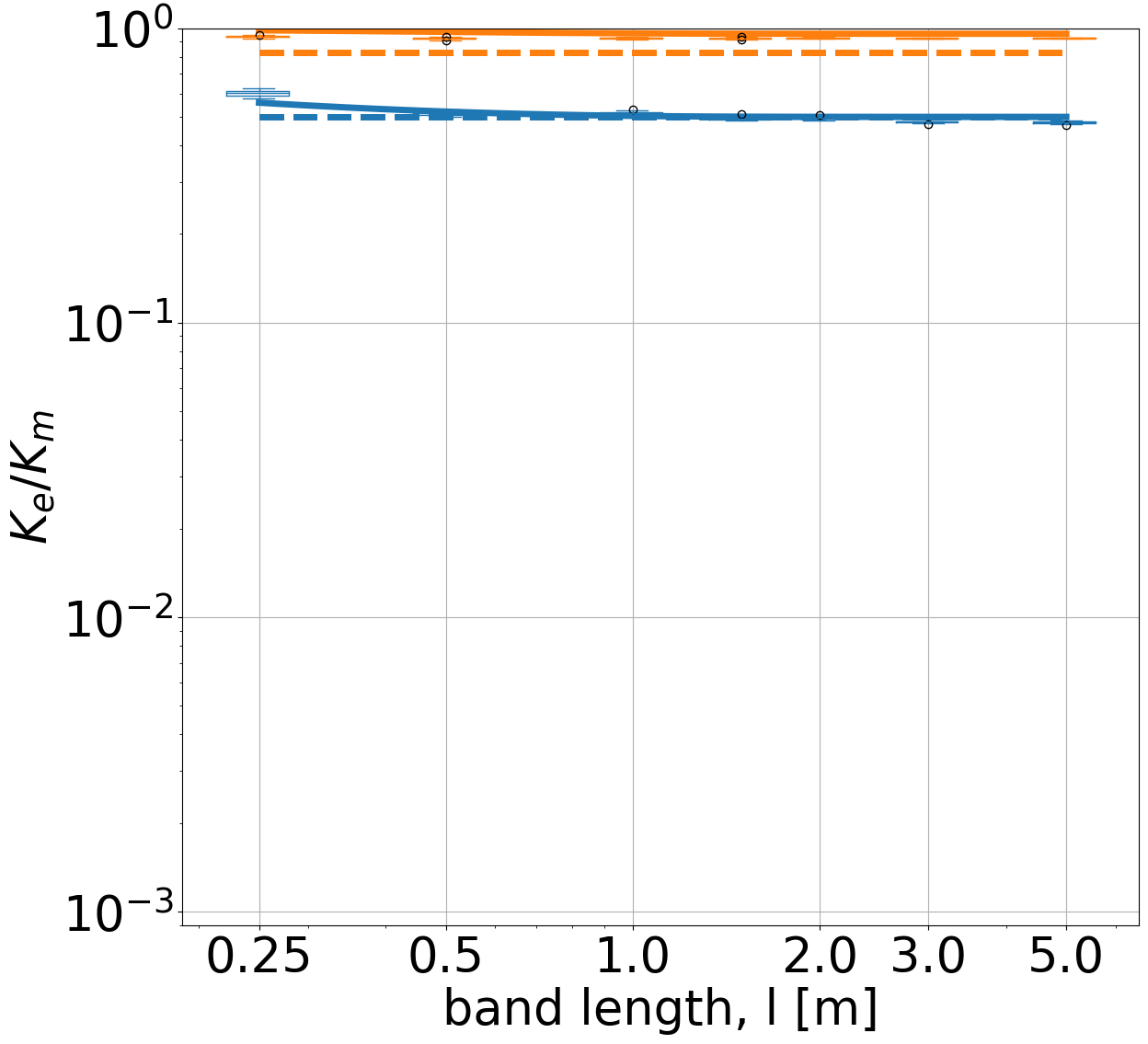}
    \caption{$\frac{K_b}{aK_m} = 10^{1}$ m$^{-1}$}
    \end{subfigure}
    \caption{Effective permeability as function of band length for the layered model (solid lines), the harmonic average (dashed lines) and the fine scale numerical simulations (bar plots). The band density is constant in the domain. The colors represent the effective permeability in $x$- and $y$-direction. The other parameters used are $\{ \rho_x, \sigma\} = \{ 10\ \text{m}^{-1}, \pi/12\}.$}
    \label{fig:K_e_Vs_band_length}
\end{figure}

Two different simulation sets are run. In the first, the band length, $l$, is varied between 0.25 m and 5 m, while the band density along a scanline is fixed to $\rho_x = 10$ m$^{-1}$. Figure~\ref{fig:K_e_Vs_band_length} shows the effective permeability of the simulation domain for the harmonic average (Equation~\eqref{eq:KeLowerBound}), the layered model (Equation~\eqref{eq:effective_perm}), and from fine-scale numerical simulations. We observe that the harmonic average consistently over-estimates the reduction in permeability due to the deformation bands, especially for smaller scaled permeability ratios $K_b / (aK_m) \le 10^{-1}$ m$^{-1}$. For higher scaled permeability ratios the harmonic average gives better results, but in these cases, the reduction of the permeability is smaller. From the figure we can also observe that as the band length increases, the effective permeability calculated by the fine scale numerical simulations approaches the harmonic average, at least for the effective permeability in the $x$-direction. The effective permeability calculated using the layered model gives results that are closer to the fine scale numerical simulations. The increase in effective permeability when the band length decreases is correctly captured in this model.

In the second set of simulations, the band length is fixed to $l=1$ m, while the band density along a scanline is varied. The effective permeability is depicted in Figure~\ref{fig:K_e_Vs_band_density}. We observe the same qualitative behaviour as when the band length is varied. For either large scaled permeability ratios, $K_b / (aK_m)\ge 10^0$ m$^{-1}$ or high band density, $\rho_x \ge 20$ m$^{-1}$, the  harmonic average give good results. However, only the layered model is able to capture the increase in permeability when the band density or band length decreases and the network becomes disconnected.

\ifpicture%
\begin{figure}
    \centering
    \newcommand{\figWidth}{0.24\textwidth}
    \includegraphics[width=0.3\textwidth]{figures/legend.png}\\
    \begin{subfigure}[b]{\figWidth}
    \includegraphics[width=\textwidth]{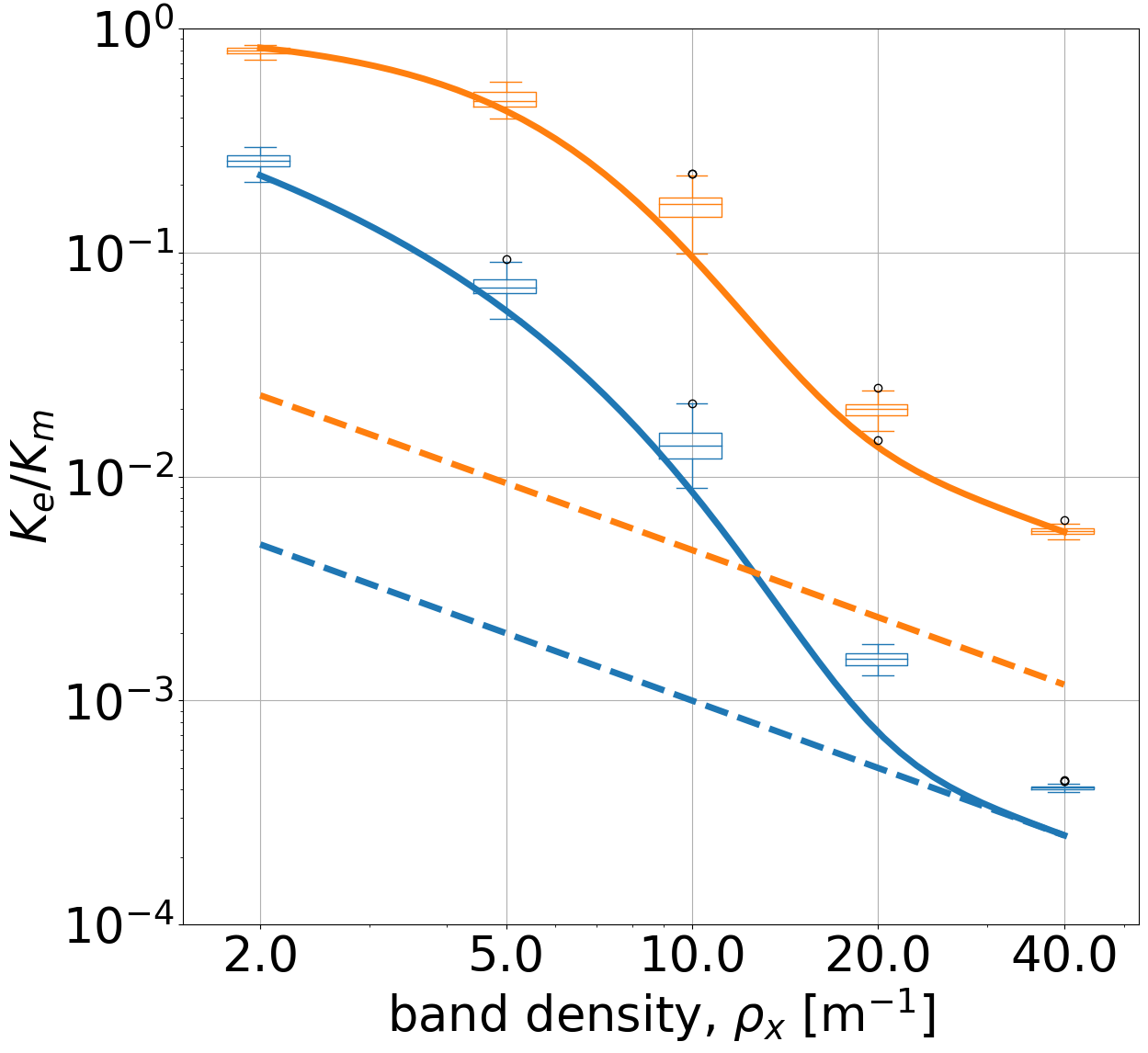}
    \caption{$\frac{K_b}{aK_m} = 10^{-2}$ m$^{-1}$}
    \end{subfigure}
    \begin{subfigure}[b]{\figWidth}
    \includegraphics[width=\textwidth]{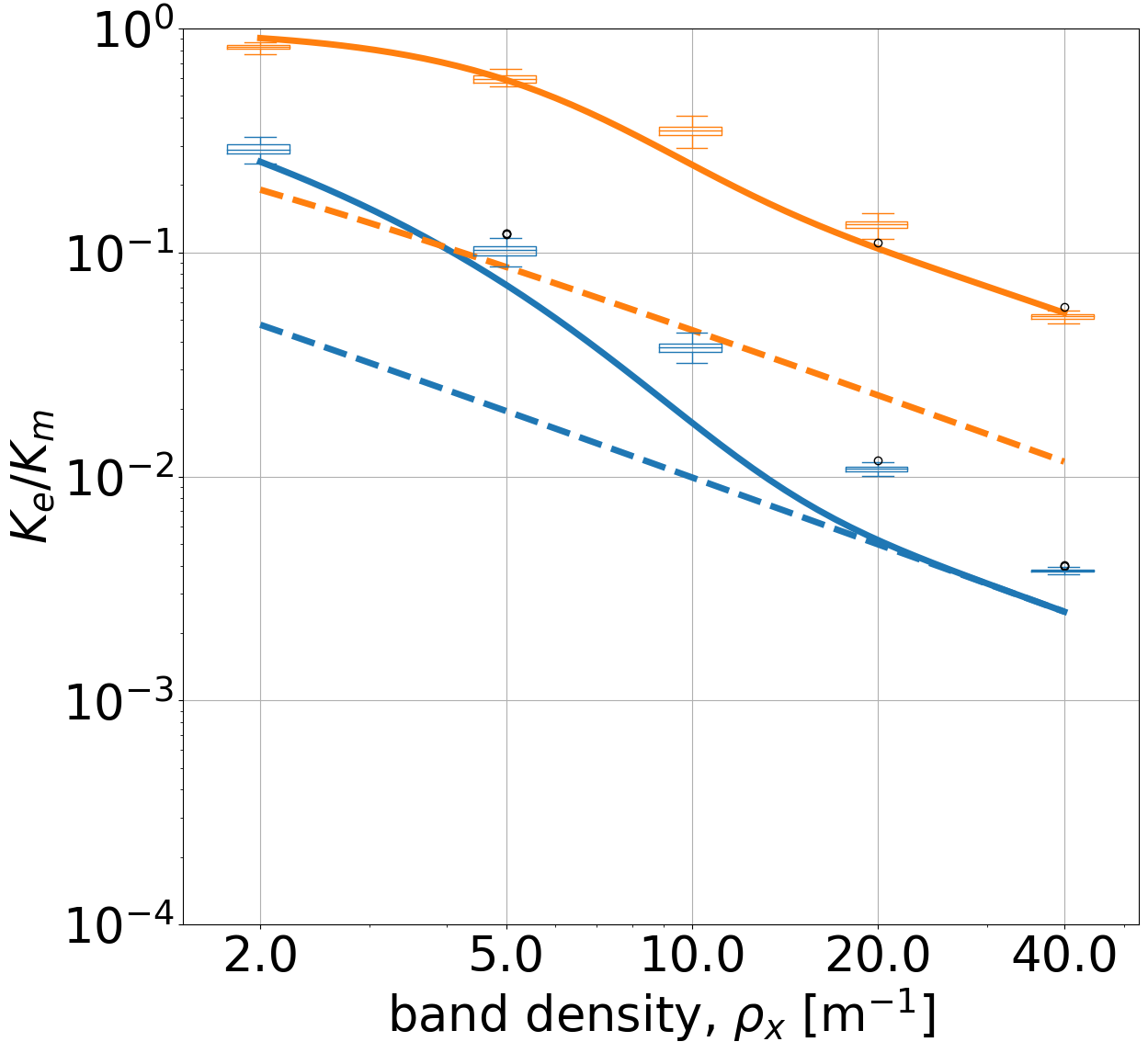}
    \caption{$\frac{K_b}{aK_m} = 10^{-1}$ m$^{-1}$}
    \end{subfigure}
    \begin{subfigure}[b]{\figWidth}
    \includegraphics[width=\textwidth]{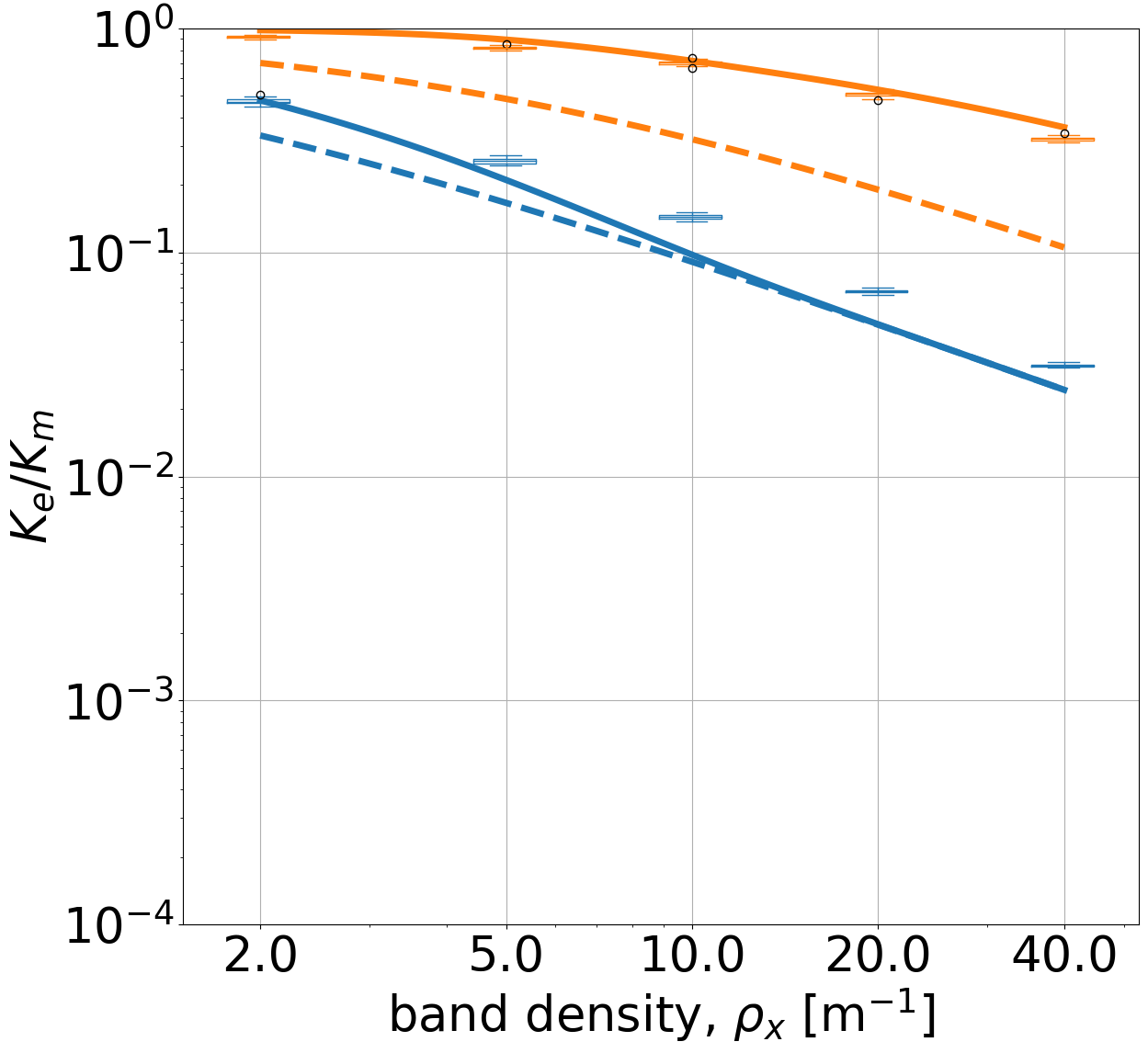}
    \caption{$\frac{K_b}{aK_m} = 10^{0}$ m$^{-1}$}
    \end{subfigure}
    \begin{subfigure}[b]{\figWidth}
    \includegraphics[width=\textwidth]{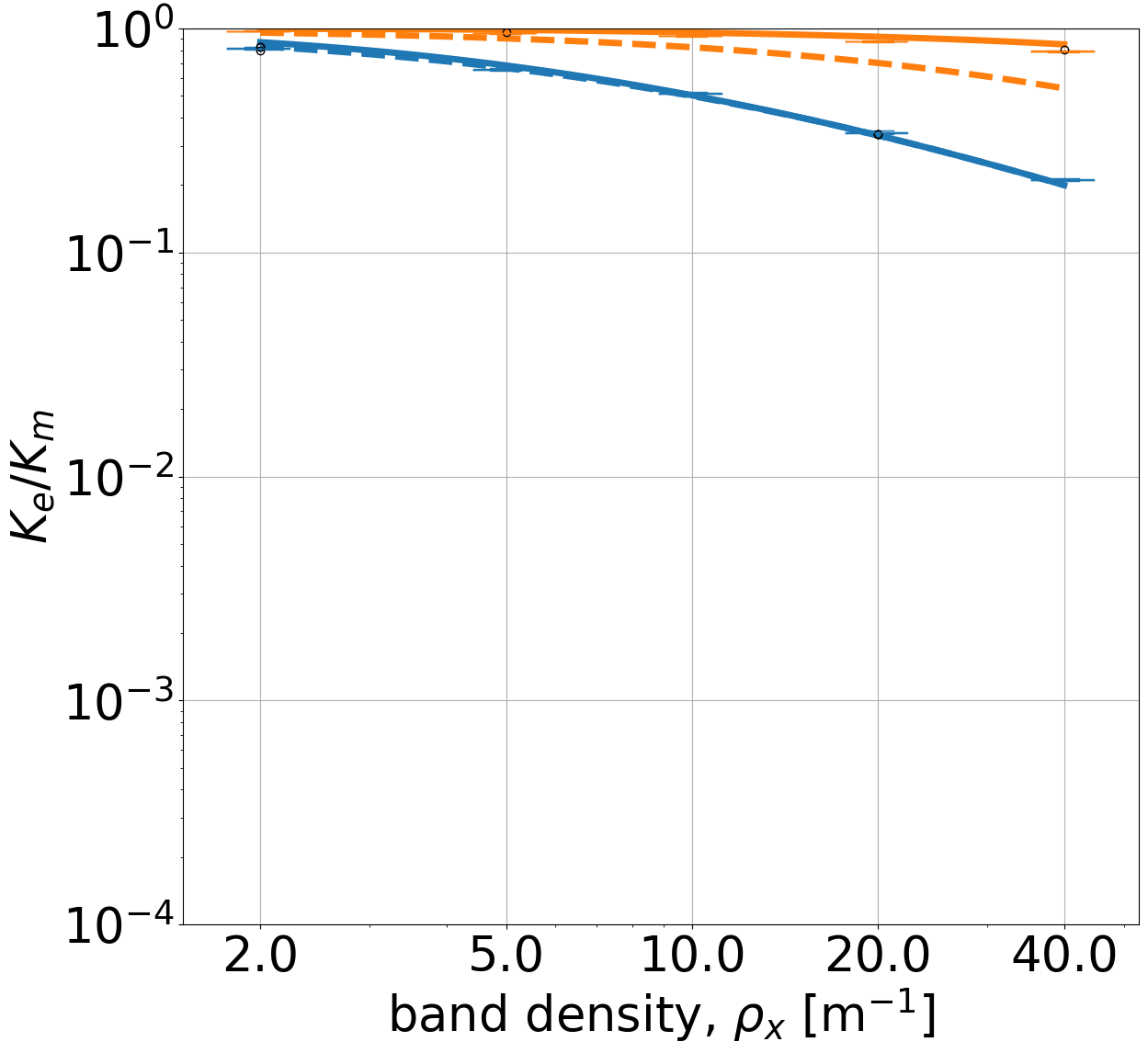}
    \caption{$\frac{K_b}{aK_m} = 10^{1}$ m$^{-1}$}
    \end{subfigure}
    \caption{Effective permeability as function of band density along a scanline for the layered model (solid lines), the harmonic average (dashed lines) and the fine scale numerical simulations (bar plots). The band density is constant in the domain. The colors represent the effective permeability in $x$- and $y$-direction. The other parameters used are $\{l, \sigma\} = \{1\ \text{m}, \pi/12\}.$}
    \label{fig:K_e_Vs_band_density}
\end{figure}
\fi

\subsection{Case with varying band density} \label{sec:res_fault}
Both the layered conceptual model given by Equation~\eqref{eq:effective_perm} and the harmonic average given by Equation~\eqref{eq:KeLowerBound} assumes constant band density. In the vicinity of faults, the band density is varying according to Equation~\eqref{eq:fault_density}, thus, we can not use the analytical solutions given in Equations~\eqref{eq:effective_perm} and~\eqref{eq:KeLowerBound} directly. However, we assume that they give appropriate point estimates of the effective permeability. 


We calculate the effective bulk permeability by dividing it into 99 equally spaced cells in the $x$-direction. Each cell is assigned a permeability based on the logarithmic density function given by Equation~\eqref{eq:fault_density} and the upscaled permeability of the layered model given by Equation~\eqref{eq:effective_perm}. The effective permeability of the domain, $K_{\alpha, e}^{dz}$, $\alpha\in\{x,y\}$,  is then calculated as the harmonic average of these cells in the $x$-direction and as an arithmetic average in the $y$-direction:
\begin{equation}\label{eq:effective_perm_variable_rho}
    \begin{aligned}
    K_{x, e}^{dz} &= \frac{L_x}{\sum_{n=1}^{99}\frac{\frac{L_x}{99}}{K_{x, e}^l (\rho(x_i))}},\\
    K_{y, e}^{dz} &= \frac{1}{L_x}\sum_{n=1}^{99}\frac{L_x}{99}K_{y, e}^l (\rho(x_i)).
    \end{aligned}
\end{equation}
Here, $L_x$ is the domain size in $x$-direction (see Equation~\eqref{eq:domain_width}), $x_i$ is the $x$-coordinate of the center of cell $i$, and $K_{\alpha, e}^l (\rho_x(x_i))$ is the permeability assigned to the cell in direction $\alpha$ (the effective permeability depends on the density as described in Equations~\eqref{eq:effective_perm_band_layer} and~\eqref{eq:effective_perm}). The permeability of the harmonic model is calculated similarly by replacing $K_{\alpha, e}^l$ by $K_{\alpha, e}^h$ in Equation~\eqref{eq:effective_perm_variable_rho}.

The effective permeability obtained by the numerical simulations is calculated by considering the pressure drop and flow rates over the domain boundary.

Three different sets of simulations are run, varying the band length, varying the standard deviation of the rotation, and varying the damage zone width. For each set of values in the three sets, 68 realizations of the deformation bands network are generated, and the effective permeability is calculated in two directions, totaling in $68\times(5 + 4+ 7)\times2=$ 2 176 runs.



\subsubsection{Effective permeability vs. band length}
In this subsection we study the sensitivity of the effective permeability on the band length. The damage zone width and the standard deviation of the band rotation are fixed to $W_5=5$ m and $\sigma = \pi / 12$. This damage zone width corresponds to a fault throw of approximately 12 m. 

The effective permeability obtained for the layered model, the harmonic average and the fine scale numerical simulations is depicted in Figure~\ref{fig:fault_perm_vs_band_length}. The layered model performs better than the harmonic average, however, it does overestimates the reduction in permeability by up to a factor two. For longer band lengths, $l\ge 4$ m, or higher scaled permeability ratios, $K_b/(aK_m)\ge 10^0$ m$^{-1}$, the harmonic average gives appropriate estimates of the effective permeability in direction normal to the fault. Compared to the case with constant band density, shown in Figure~\ref{fig:K_e_Vs_band_length} the shape of the effective permeability as a function of band length is different. This is captured in the layered model.

\ifpicture
\begin{figure}
    \centering
    \newcommand{\figWidth}{0.24\textwidth}
    \includegraphics[width=0.3\textwidth]{figures/legend.png}\\
    \begin{subfigure}[b]{\figWidth}
    \includegraphics[width=\textwidth]{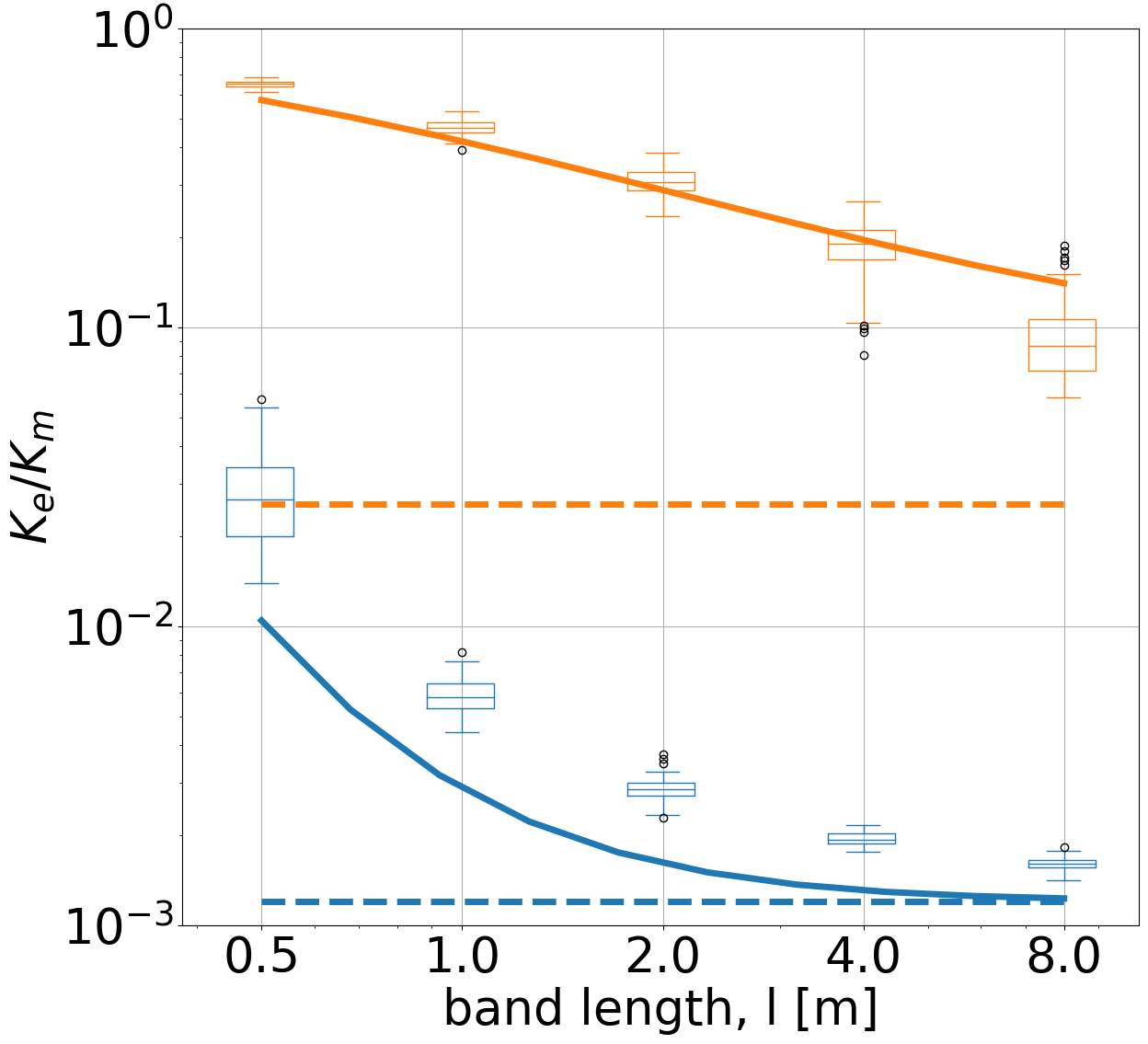}
    \caption{$\frac{K_b}{aK_m} = 10^{-2}$ m$^{-1}$}
    \end{subfigure}
    \begin{subfigure}[b]{\figWidth}
    \includegraphics[width=\textwidth]{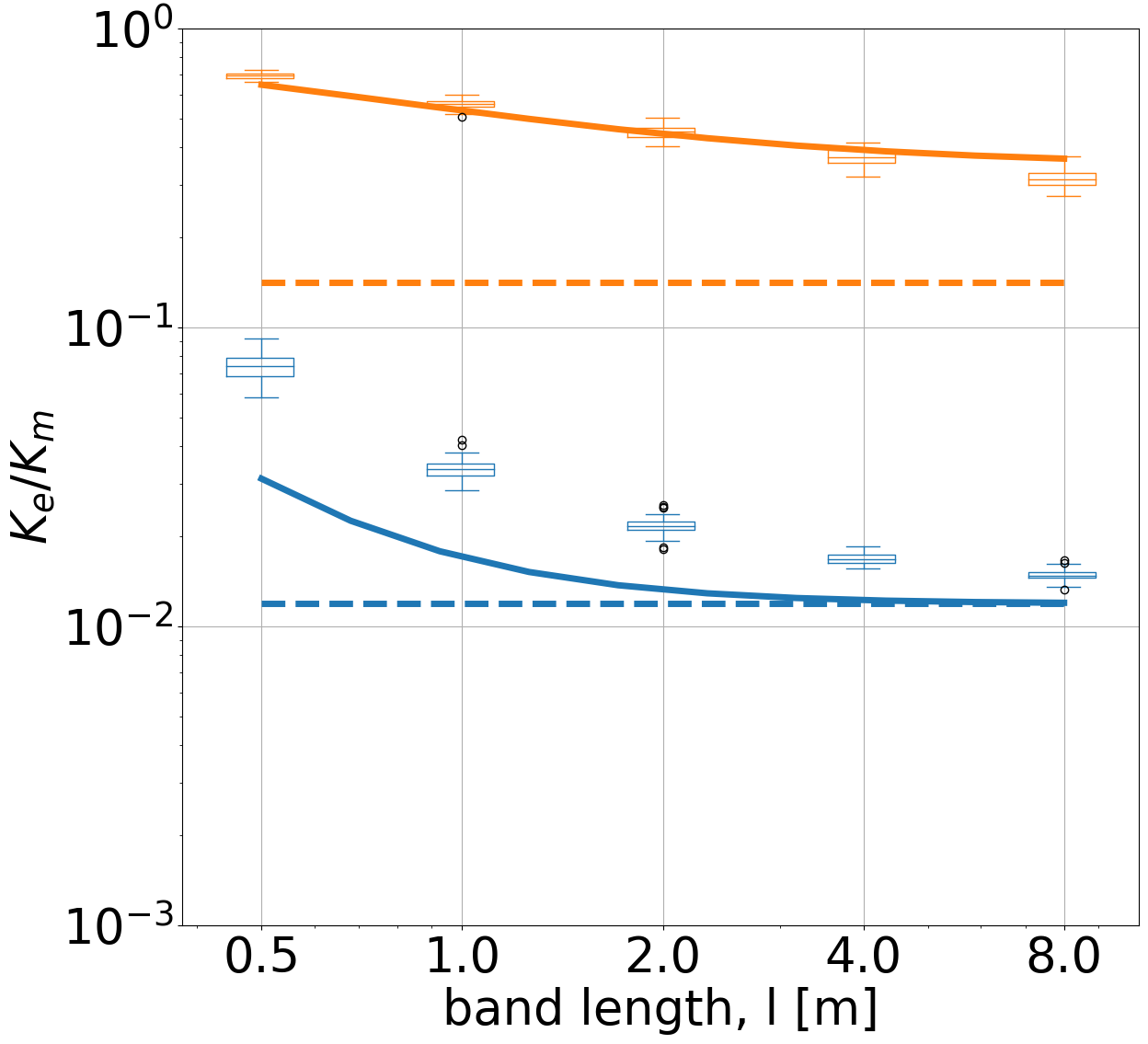}
    \caption{$\frac{K_b}{aK_m} = 10^{-1}$ m$^{-1}$}
    \end{subfigure}
    \begin{subfigure}[b]{\figWidth}
    \includegraphics[width=\textwidth]{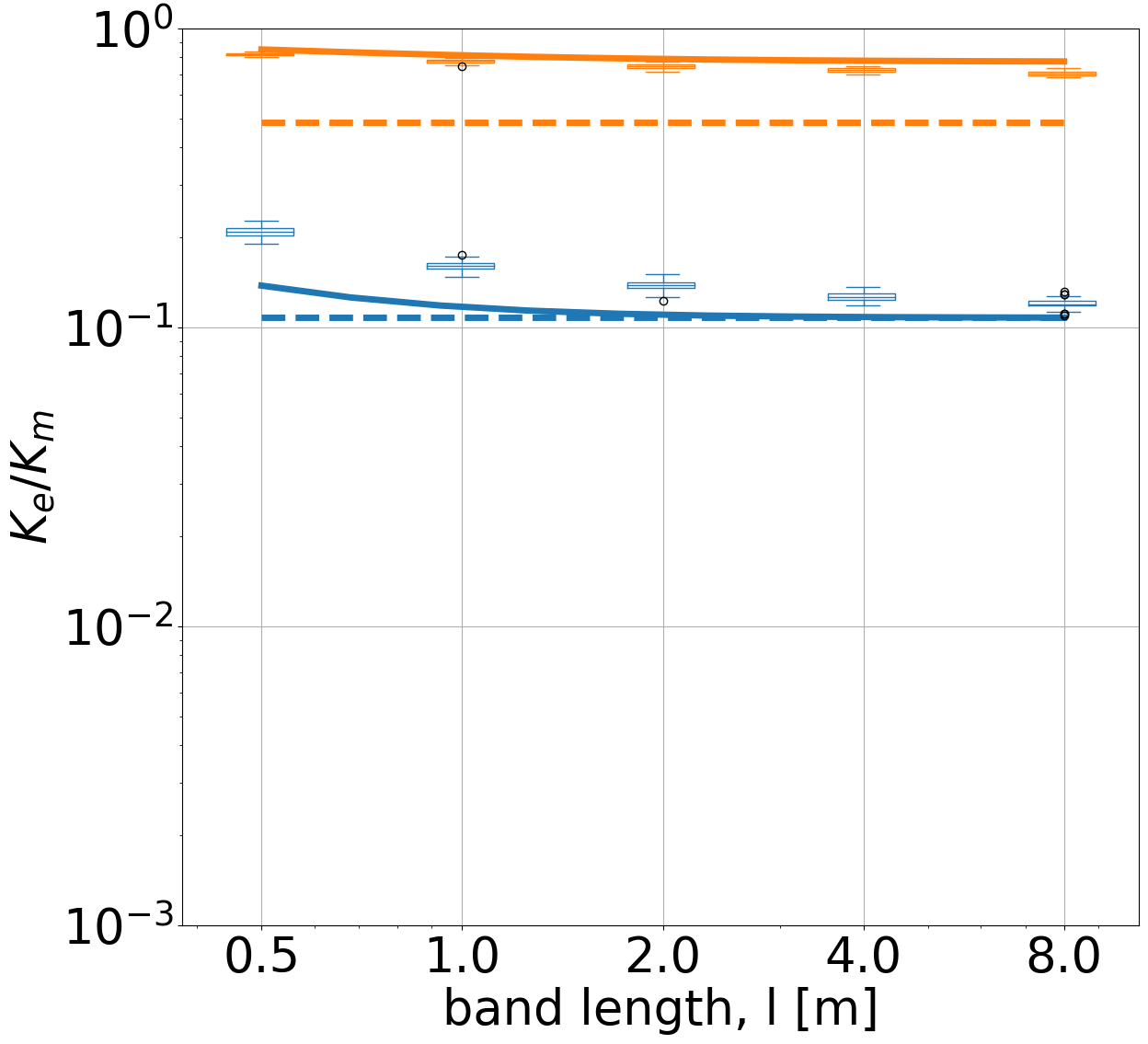}
    \caption{$\frac{K_b}{aK_m} = 10^{0}$ m$^{-1}$}
    \end{subfigure}
    \begin{subfigure}[b]{\figWidth}
    \includegraphics[width=\textwidth]{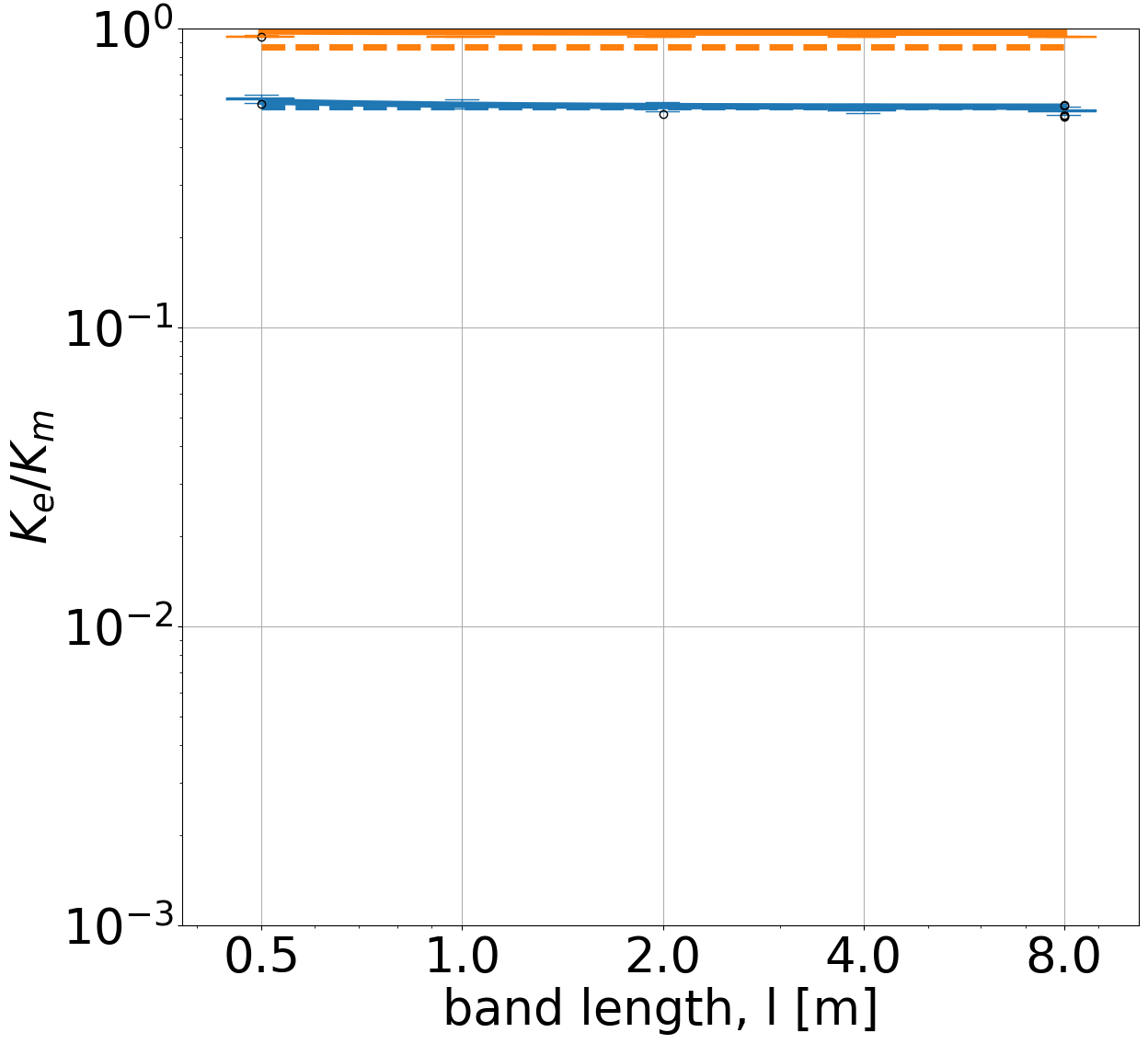}
    \caption{$\frac{K_b}{aK_m} = 10^{1}$ m$^{-1}$}
    \end{subfigure}
    \caption{Effective permeability as function of band length for the layered model (solid lines), the harmonic average (dashed lines) and the fine scale numerical simulations (bar plots). The band density follows the logarithmic function in Equation~\eqref{eq:fault_density}. The colors represent the effective permeability normal (blue) and parallel (yellow) to the fault. The other parameters used are $\{W_5, \sigma\} = \{5\ \text{m}, \pi/12\}.$}
    \label{fig:fault_perm_vs_band_length}
\end{figure}
\fi

\subsubsection{Effective permeability vs rotation}
In this subsection, the standard deviation of the rotation, $\sigma$, is varied while the band length and damage zone width are fixed to $l= 1$ m and $W_5 = 5$ m, respectively. 

The effective permeability as a function of the standard deviation of the rotation is shown in Figure~\ref{fig:fault_perm_vs_band_rot}. We observe that in the direction normal to the fault the effective permeability is only depending on $\sigma$ for smaller scaled permeability ratios, $K_b/(aK_m) \le 10^{-1}$ m$^{-1}$. For small standard deviation, the deformation bands are almost parallel to the fault, with very few intersections. As the standard deviation increases the network consist of more intersections and larger clusters, however, this only affects the effective permeability when the scaled permeability ratio is small, $K_b/(aK_m) \le 10^{-1}$ m$^{-1}$. For higher scaled permeability ratios, the resistance to flow across the deformation bands is not large enough to favour these additional pathways around the deformation bands. Thus, we can observe that the harmonic average gives good results in this case. On the other hand, the layered model is able to capture the increase in effective permeability for the smaller scaled permeability ratios and smaller standard deviation.

It is worth noting that the difference in permeability between $x$- and $y$-direction for the large values of $\sigma$ is mainly due to that the density varies in $x$-direction (and is constant i $y$-direction), which gives a region with very few deformation bands far from the fault; see Figure~\ref{fig:fault_def_band}

\ifpicture%
\begin{figure}
    \centering
    \newcommand{\figWidth}{0.24\textwidth}
    \includegraphics[width=0.3\textwidth]{figures/legend.png}\\
    \begin{subfigure}[b]{\figWidth}
    \includegraphics[width=\textwidth]{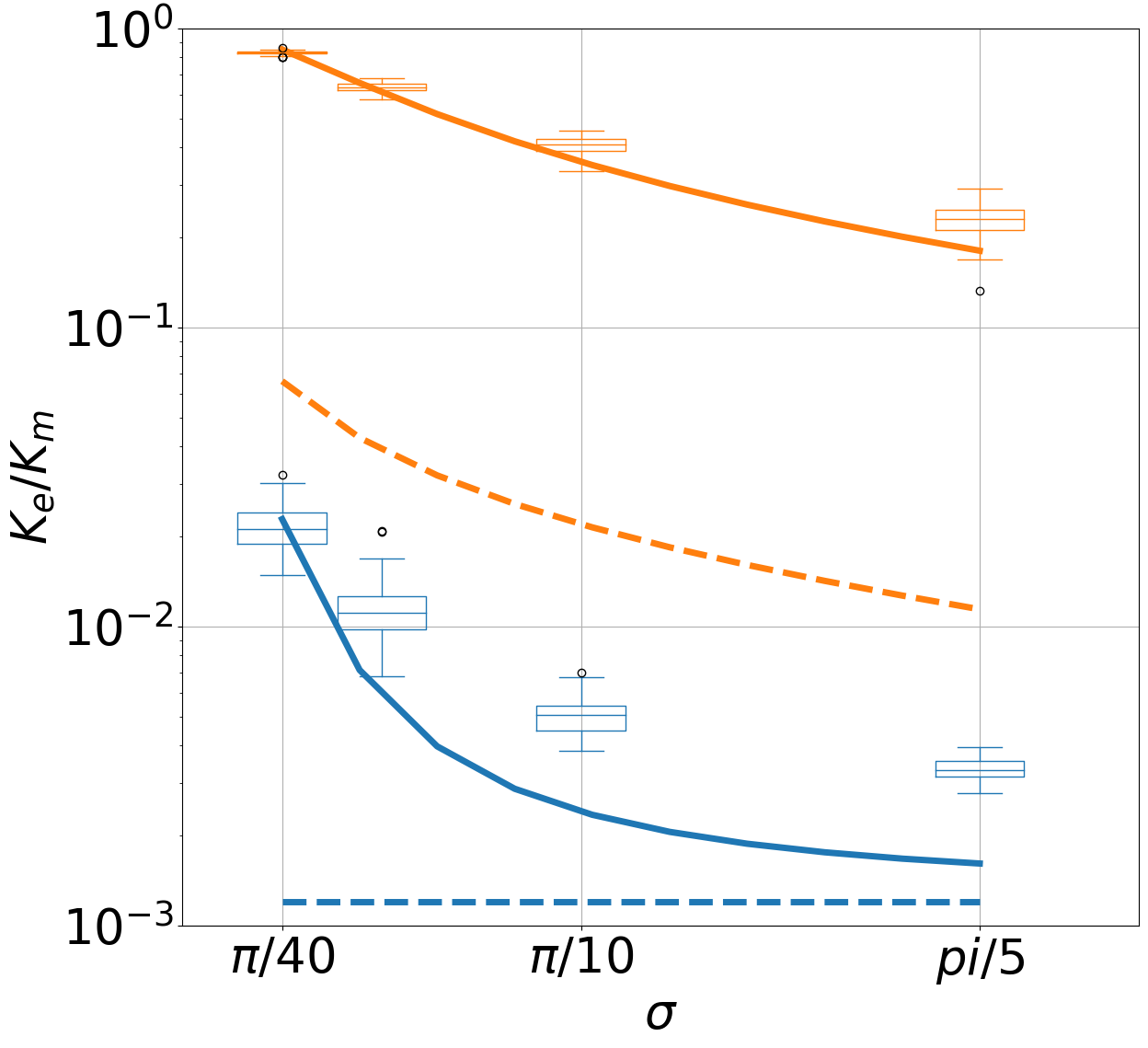}
    \caption{$\frac{K_b}{aK_m} = 10^{-2}$ m$^{-1}$}
    \end{subfigure}
    \begin{subfigure}[b]{\figWidth}
    \includegraphics[width=\textwidth]{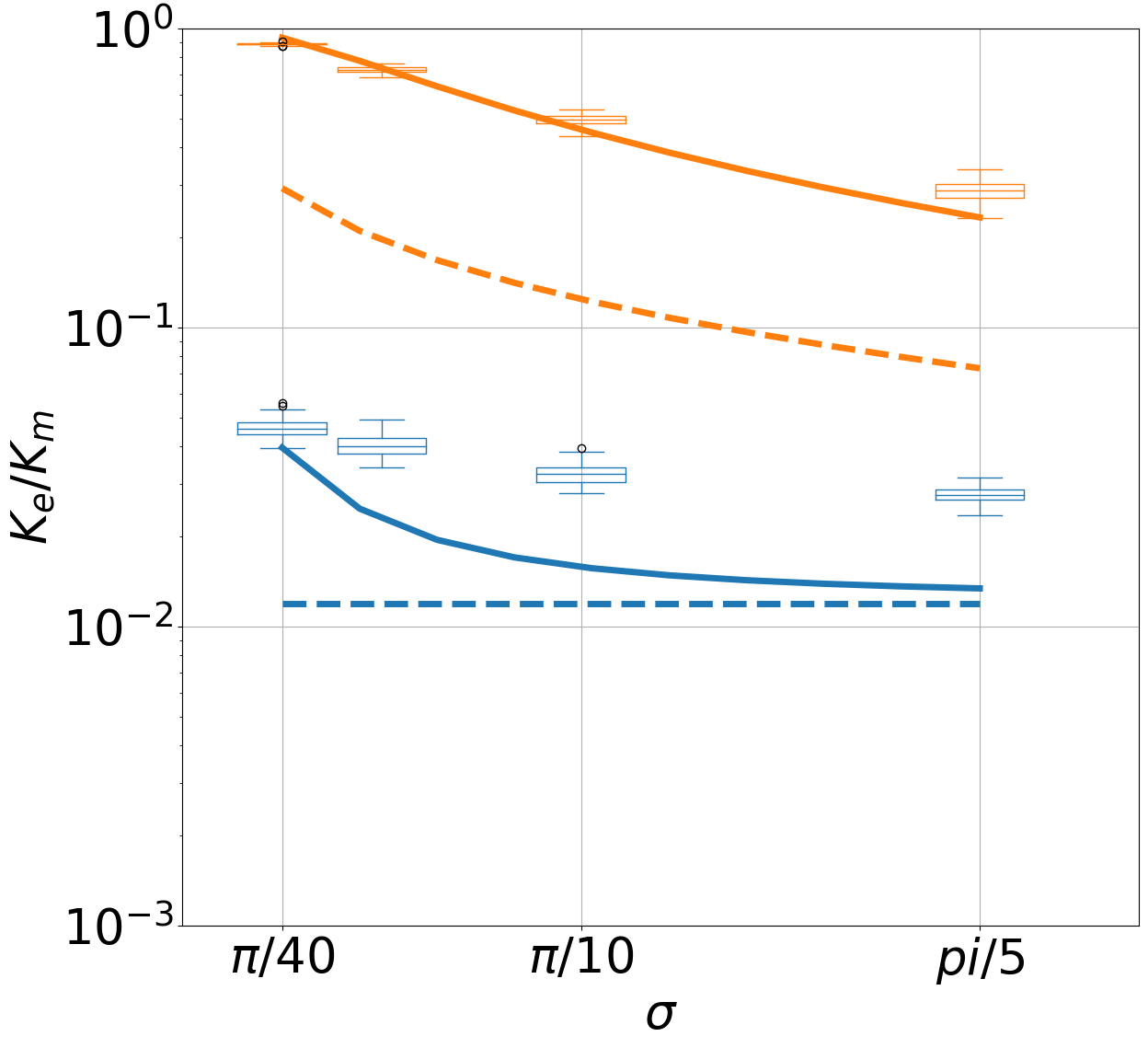}
    \caption{$\frac{K_b}{aK_m} = 10^{-1}$ m$^{-1}$}
    \end{subfigure}
    \begin{subfigure}[b]{\figWidth}
    \includegraphics[width=\textwidth]{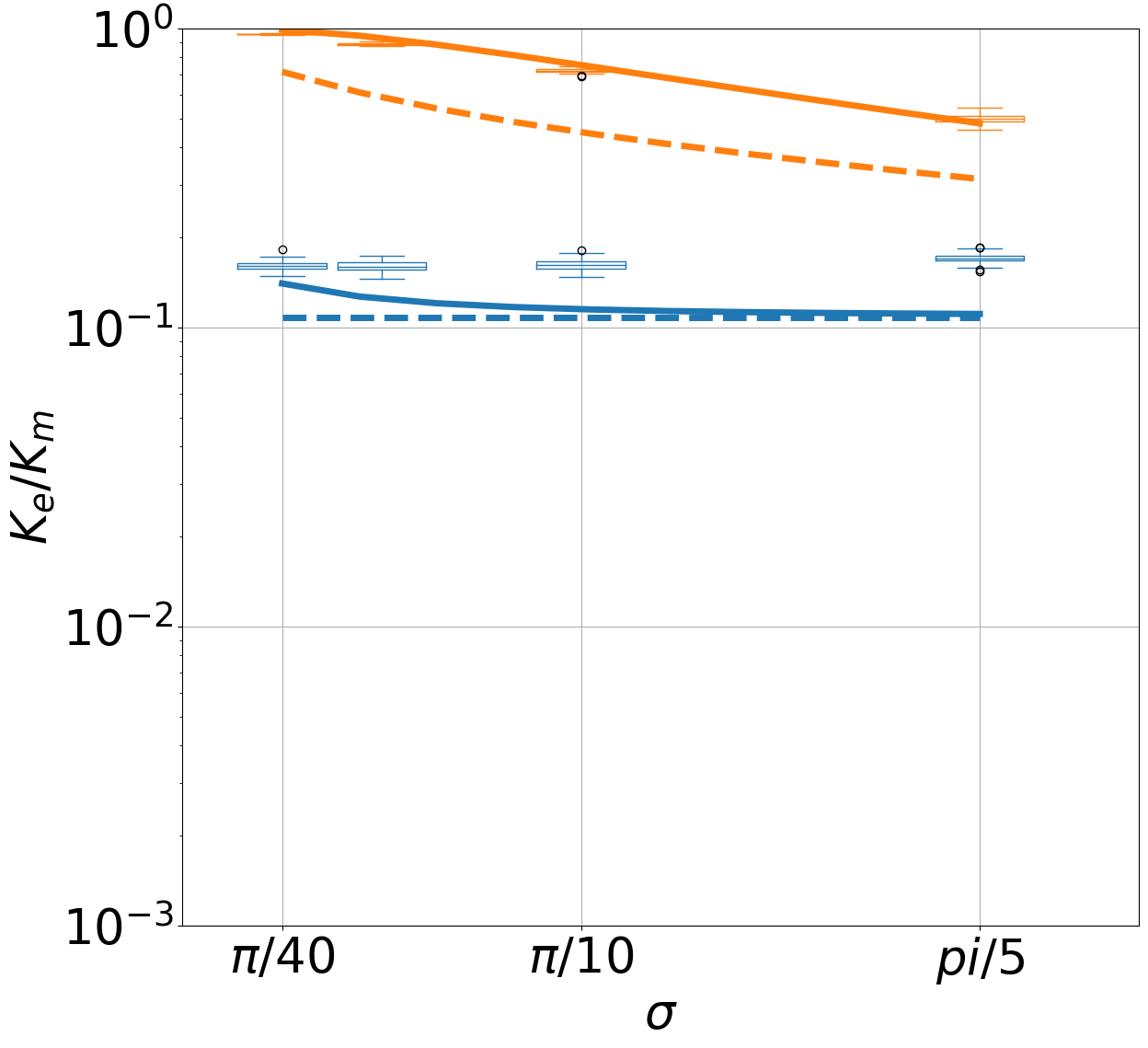}
    \caption{$\frac{K_b}{aK_m} = 10^{0}$ m$^{-1}$}
    \end{subfigure}
    \begin{subfigure}[b]{\figWidth}
    \includegraphics[width=\textwidth]{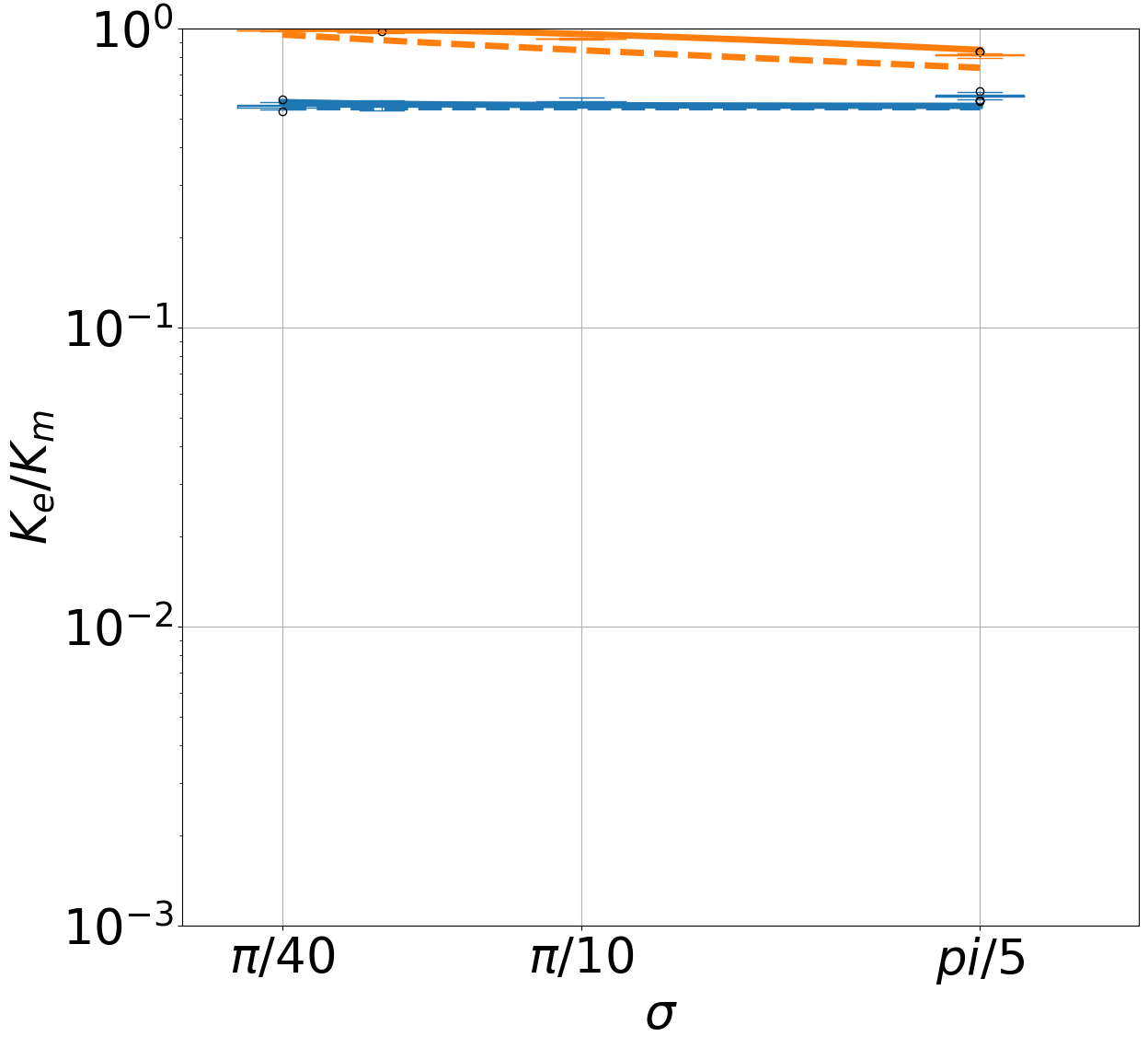}
    \caption{$\frac{K_b}{aK_m} = 10^{1}$ m$^{-1}$}
    \end{subfigure}
    \caption{Effective permeability as function of band rotation for the layered model (solid lines), the harmonic average (dashed lines) and the fine scale numerical simulations (bar plots). The band density follows the logarithmic function in Equation~\eqref{eq:fault_density}. The colors represent the effective permeability normal (blue) and parallel (orange) to the fault. The other parameters used are $\{l, W_5\} = \{1\ \text{m}, 5\ \text{m}\}.$}
    \label{fig:fault_perm_vs_band_rot}
\end{figure}
\fi

\subsubsection{Effective permeability vs damage zone width}
In this subsection, the band length and the standard deviation of the rotation is fixed to $l = 1$ m and $\sigma = \pi / 12$. The damage zone width is varied between $W_5 = $ 1 m, 2 m, 3 m, 5 m, 7 m, 10 m, and 20 m, which by using the relation in Equation~\eqref{eq:widthFromThrow} correspond to fault throws of 0.28 m, 1.4 m, 3.5 m, 12 m, 25 m, 58 m and 293 m.

Figure~\ref{fig:fault_perm_vs_W5} shows the effective permeability as a function of damage zone width. For all scaled permeability ratios the effective permeability have the same qualitative behaviour. The effective permeability is not depending on the damage zone width, and we observe a larger variation between different realizations of the deformation band network for smaller damage zone widths. The reason that the effective permeability is independent of damage zone width is the assumption on the density profile in Equation~\eqref{eq:fault_density}: The average density of deformation bands in the damage zone is assumed independent of throw. However, we note that it is only the average effective permeability of the simulation domain that is independent of damage zone width. Both the layered model and the harmonic average capture this qualitative behaviour, but the layered model gives better estimates of the effective permeability, especially in the direction parallel to the fault.
\ifpicture%
\begin{figure}
    \centering
    \newcommand{\figWidth}{0.24\textwidth}
    \includegraphics[width=0.3\textwidth]{figures/legend.png}\\
    \begin{subfigure}[b]{\figWidth}
    \includegraphics[width=\textwidth]{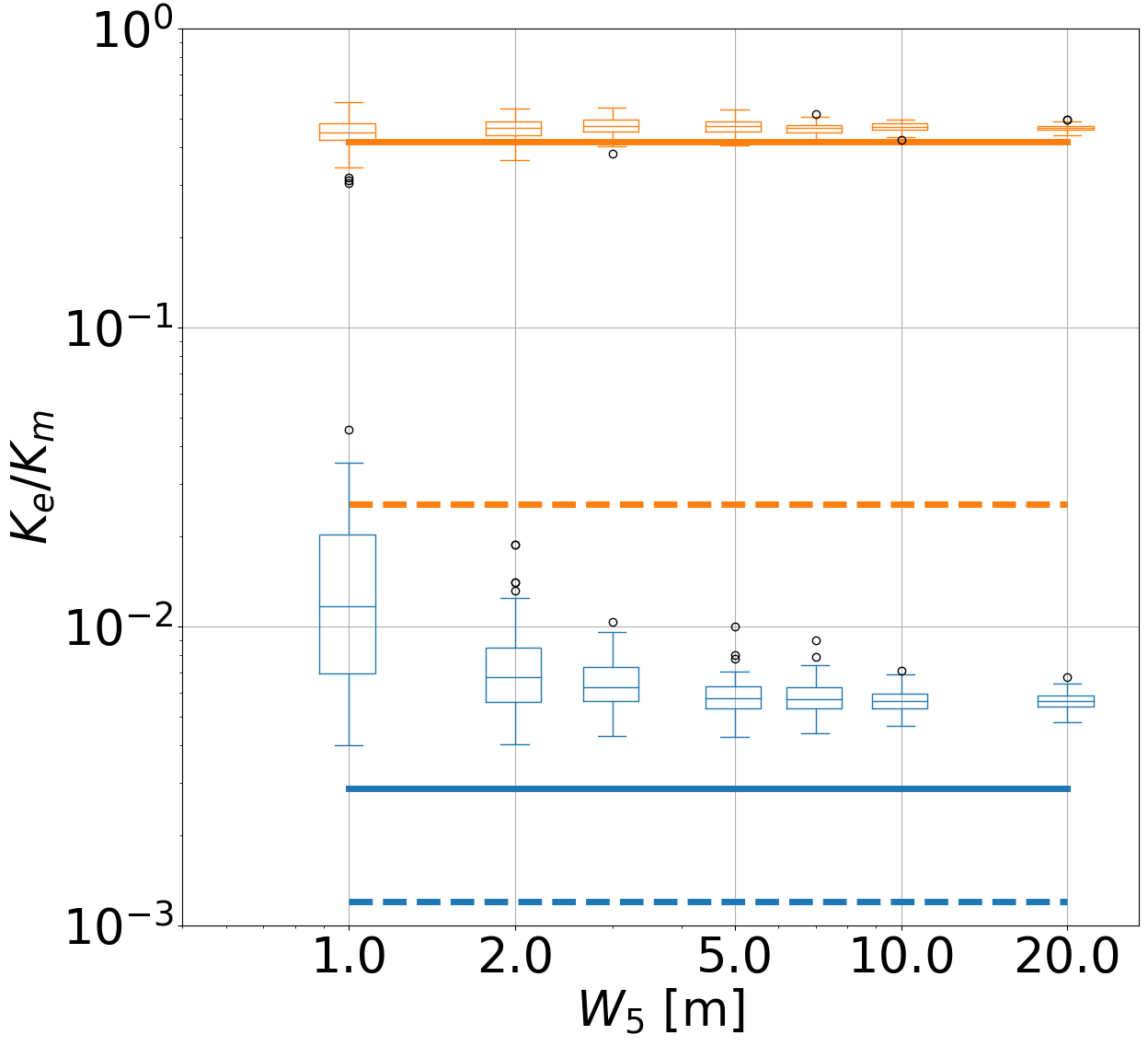}
    \caption{$\frac{K_b}{aK_m} = 10^{-2}$ m$^{-1}$}
    \end{subfigure}
    \begin{subfigure}[b]{\figWidth}
    \includegraphics[width=\textwidth]{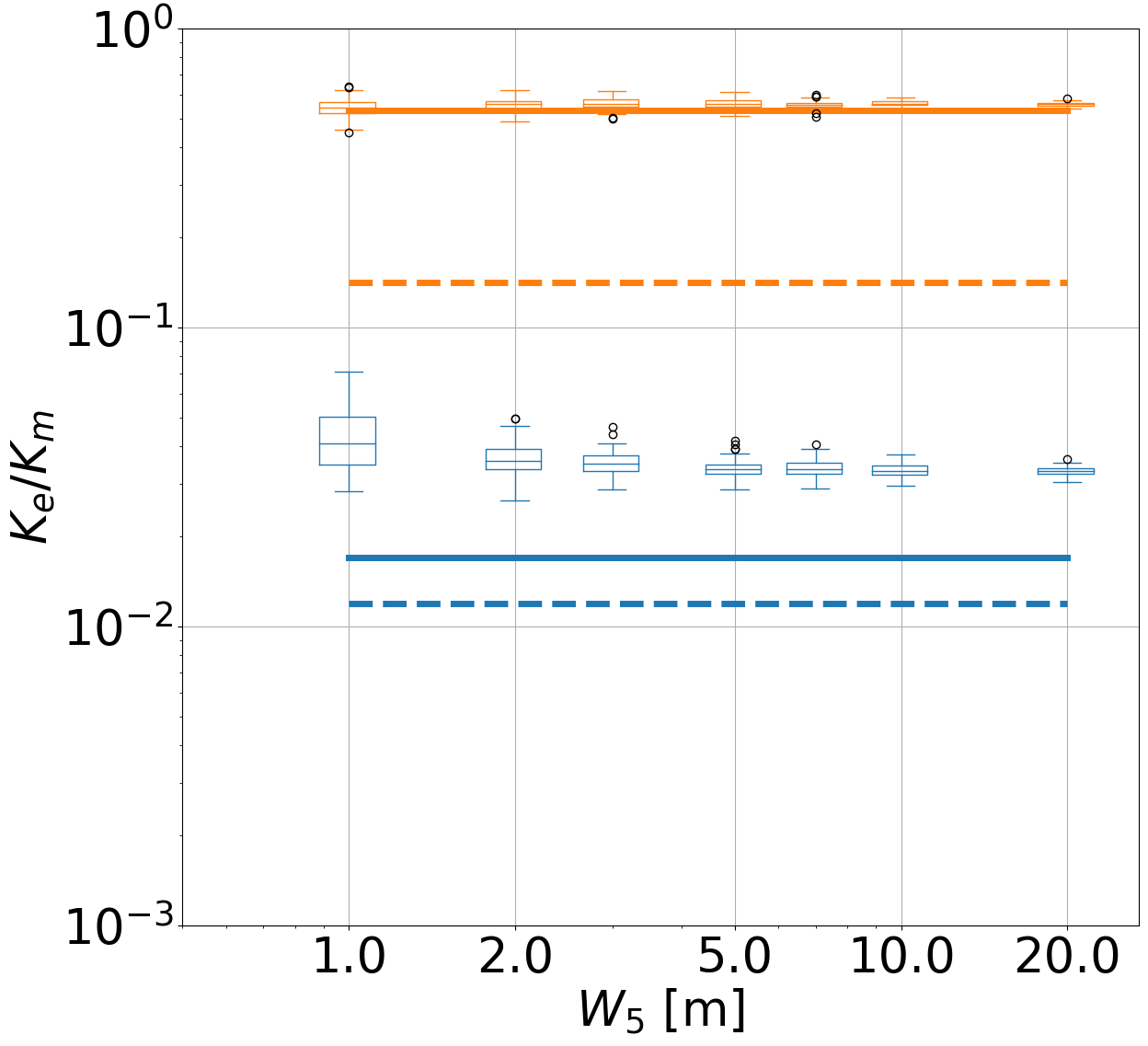}
    \caption{$\frac{K_b}{aK_m} = 10^{-1}$ m$^{-1}$}
    \end{subfigure}
    \begin{subfigure}[b]{\figWidth}
    \includegraphics[width=\textwidth]{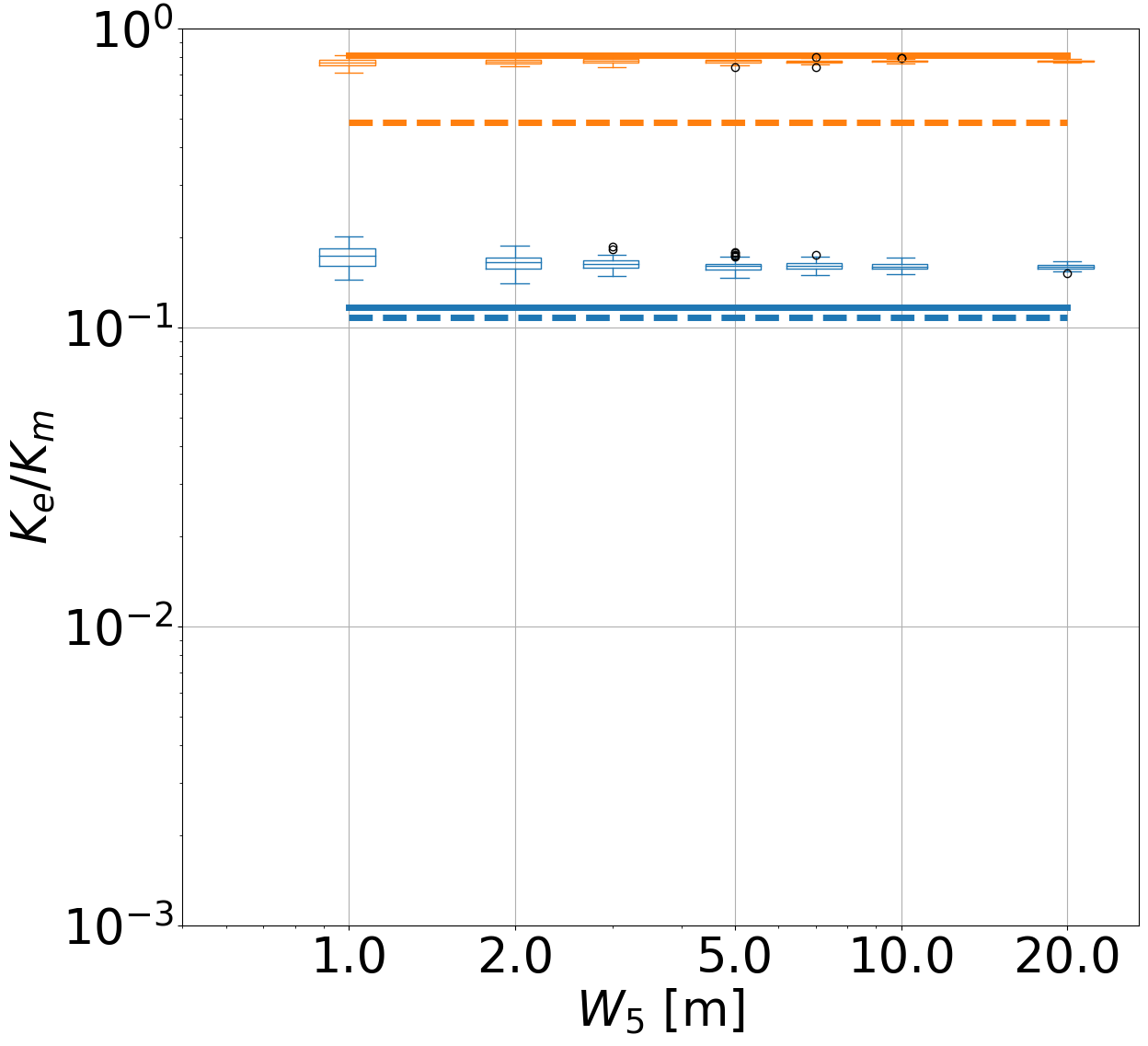}
    \caption{$\frac{K_b}{aK_m} = 10^{0}$ m$^{-1}$}
    \end{subfigure}
    \begin{subfigure}[b]{\figWidth}
    \includegraphics[width=\textwidth]{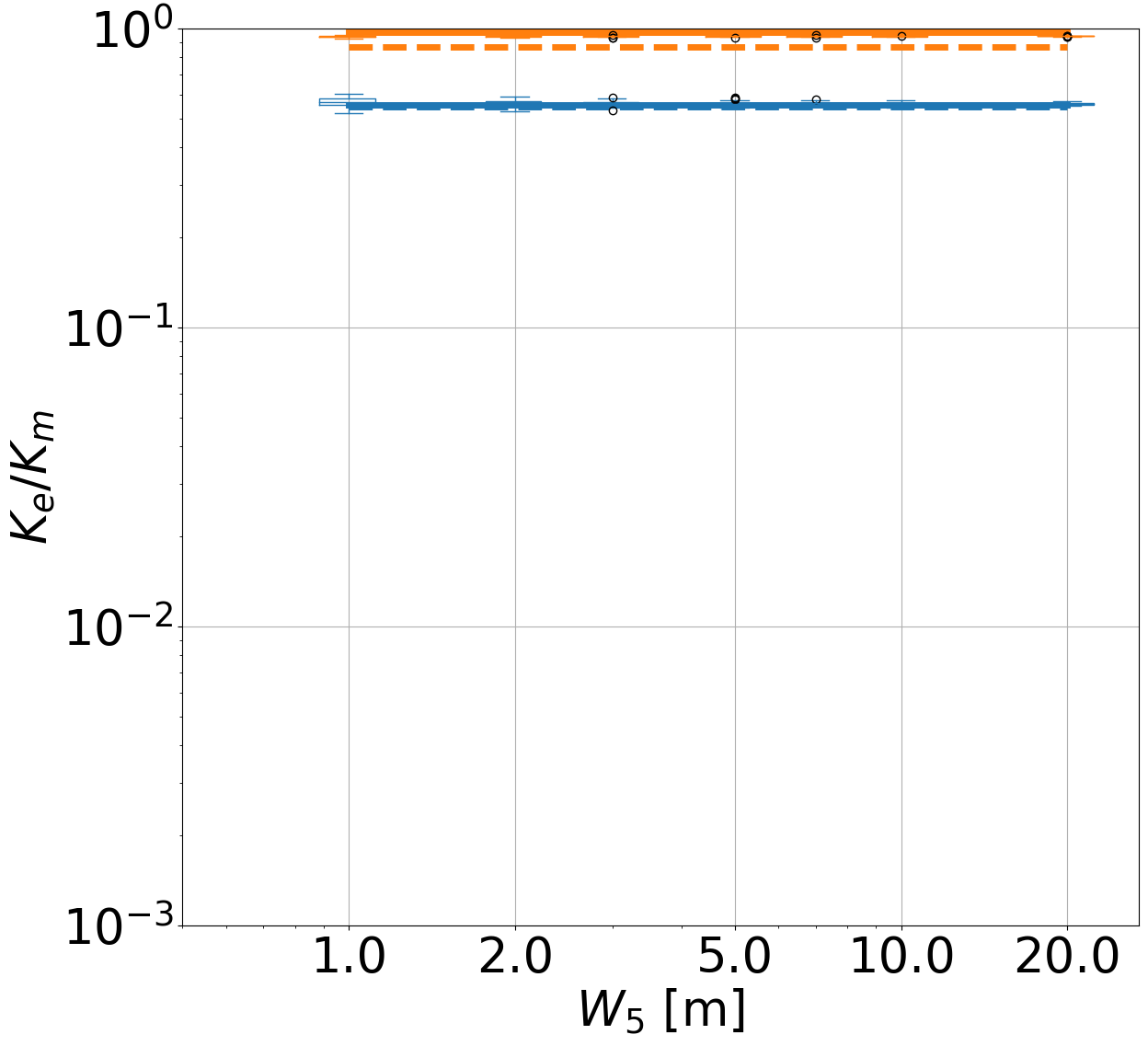}
    \caption{$\frac{K_b}{aK_m} = 10^{1}$ m$^{-1}$}
    \end{subfigure}
    \caption{Effective permeability as function of damage zone width for the layered model (solid lines), the harmonic average (dashed lines) and the fine scale numerical simulations (bar plots). The band density follows the logarithmic function in Equation~\eqref{eq:fault_density}. The colors represent the effective permeability normal (blue) and parallel (orange) to the fault. The other parameters used are $\{l, \sigma\} = \{1\ \text{m}, \pi/12\}.$}
    \label{fig:fault_perm_vs_W5}
\end{figure}
\fi%
\noindent

\section{Incorporating deformation bands in field-scale simulations}
\label{sec:deformation_bandsfield_scale}
The analytical expressions derived in the previous sections can be used to incorporate the effect of deformation bands into field-scale simulation models. In Section~\ref{sec:field_scale_sim} we demonstrate this by applying the model to a \co storage scenario where we study the effect of deformation bands on leakage along faults. In this section we discuss how to properly include the effective permeability of the damage zone calculated by Equation~\eqref{eq:effective_perm_variable_rho} into standard reservoir simulators.

The first challenge is to identify a practical approach to simulate along-fault fluid flow. 
Conceptually two types of approaches can be used to model the leakage along faults, which are distinguished by their geometric representation of the fault and the connected aquifers, as well as the fluid flow between these. 
To study fluid flow through the fault in detail, the fault must be explicitly represented in the simulation model. 
This is commonly achieved by considering the fault as a geometric object separate from the main reservoir, where the permeability normal and tangential to the fault control flow into and along the fault, respectively, see for instance \cite{martin_modeling_2005, starnoni2021}.
In practice, such models can be implemented by the coupling of different geometric domains \cite{koch2021dumux,keilegavlen2021porepy}. This approach, however, needs specialized simulators tailored to solve multi-domain problems.

An alternative and significantly simpler approach that is more aligned with the standard reservoir engineering workflow is to use numerical aquifers that are connected to the simulation grid by non-neighboring connections (NNCs). Numerical aquifers in reservoir simulation represent the hydraulically connected pore volume that exists outside of the simulation domain.  Normally a numerical aquifer represents the lateral extension of the reservoir, however, there are no formal requirements to its location. As such, a numerical aquifer can be used to represent an overlying aquifer connected hydraulically to the reservoir via a fault. 
The NNC feature is used to connect the overlying numerical aquifer to reservoir cells that lie along the fault boundary. The correct flow properties of the fault and associated damage zone can be captured in the transmissibility assigned to the NNC connections.
While this approach does not resolve the fluid flow within the fault system and thus cannot be used to study the leakage in detail, it will represent the larger-scale migration of fluid up to and potentially through the fault.

The second challenge to overcome is to handle the discrepancy of length scales between the damage zone and the cell size in the numerical model. If the computational grid is sufficiently refined, the effective permeability of the deformation bands calculated by Equation~\eqref{eq:effective_perm} can be assigned directly to the cells in the main storage formation. 
However, the cell size in the storage formation may be significantly larger than the width of the damage zone; for instance the computational grid used in Section~\ref{sec:field_scale_sim} is roughly $400\times 400$~m$^2$ in lateral extension, while the damage zone calculated from Equation~\eqref{eq:widthFromThrow} is of length $\sim{}25$~m. This causes there to be a substantial permeability heterogeneity within simulation cells close to the fault which will give large numerical errors. An alternative is to incorporate the effect of the deformation bands as a multiplier between the main formation and the fault. 
This is the approach used in Section~\ref{sec:field_scale_sim}.

To calculate the multiplier, we estimate the permeability reduction from reservoir to the fault core due to the deformation bands as
\begin{equation}
M_{db} = \frac{K_r}{  K_r(1 - D) + D}, \quad D = X / D_{cell}, \quad K_r = K_{x, e}^{dz} / K_{cell}.
\label{eq:db_mult}
\end{equation}
Here, $X = \exp(-A/B)$ is the distance from the fault at which the band density is zero (see Equation~\eqref{eq:fault_density}), and $D_{cell}$ is the distance from the cell center to its face connected to the fault.
The permeability $K_{x, e}^{dz}$ represents the effective permeability of the damage zone normal to the fault using the layered model (see Equation~\eqref{eq:effective_perm_variable_rho}), while $K_{cell}$ represents the permeability of the host cell.
Now, the transmissibility used for the NNCs from the reservoir to the aquifer is given by the harmonic mean of the half transmissibility computed from the reservoir side, $T_{r}$, multiplied by the effect of the deformation bands, $M_{db}$, and the transmissibility computed from the aquifer side, $T_{aq}$, as
\begin{equation}\label{eq:num_aqf_trans}
T_{r-aq} = \frac{M_{db} T_{r} T_{aq}}{M_{db} T_{r} + T_{aq}},  \quad T_{aq} = K_{f} A_{f} / L_{f}. 
\end{equation}
The fault area, $A_f$ represents the area available for flow between the reservoir cell and the aquifer, and the length, $L_f$, represents the length between the reservoir and aquifer.

\section{Application: Field-scale simulation}\label{sec:field_scale_sim}
While the fine-scale simulations presented in Section \ref{sec:validation} show that the deformation bands can have a significant impact on the local permeability in a fault damage zone, the effect on \co storage operations depends on fluid flow in the wider injection formation.
A full assessment of this flow, including possible leakage through the fault, would require representation of two-phase effects caused by the deformation bands, which is beyond the scope of this work.
Still, to indicate the effects deformation bands can have on flow transport through a fault in a \co injection scenario, we set up a simulation of a large-scale \co injection into the Smeaheia formation in the North Sea, which has the Vette fault as one of its structural boundaries~\cite{mulrooney2020}.
The goals of the simulation are first to give a proof of concept that deformation bands can be included in field-scale simulations with relatively minor modifications of industry standard simulation setups, and second to gain insight into the potential impact of deformation bands on fluid flow through the fault.

\subsection{Simulation setup}
\label{sec:sim_setup}
\begin{figure}
    \centering
    \includegraphics[width=0.8\textwidth]{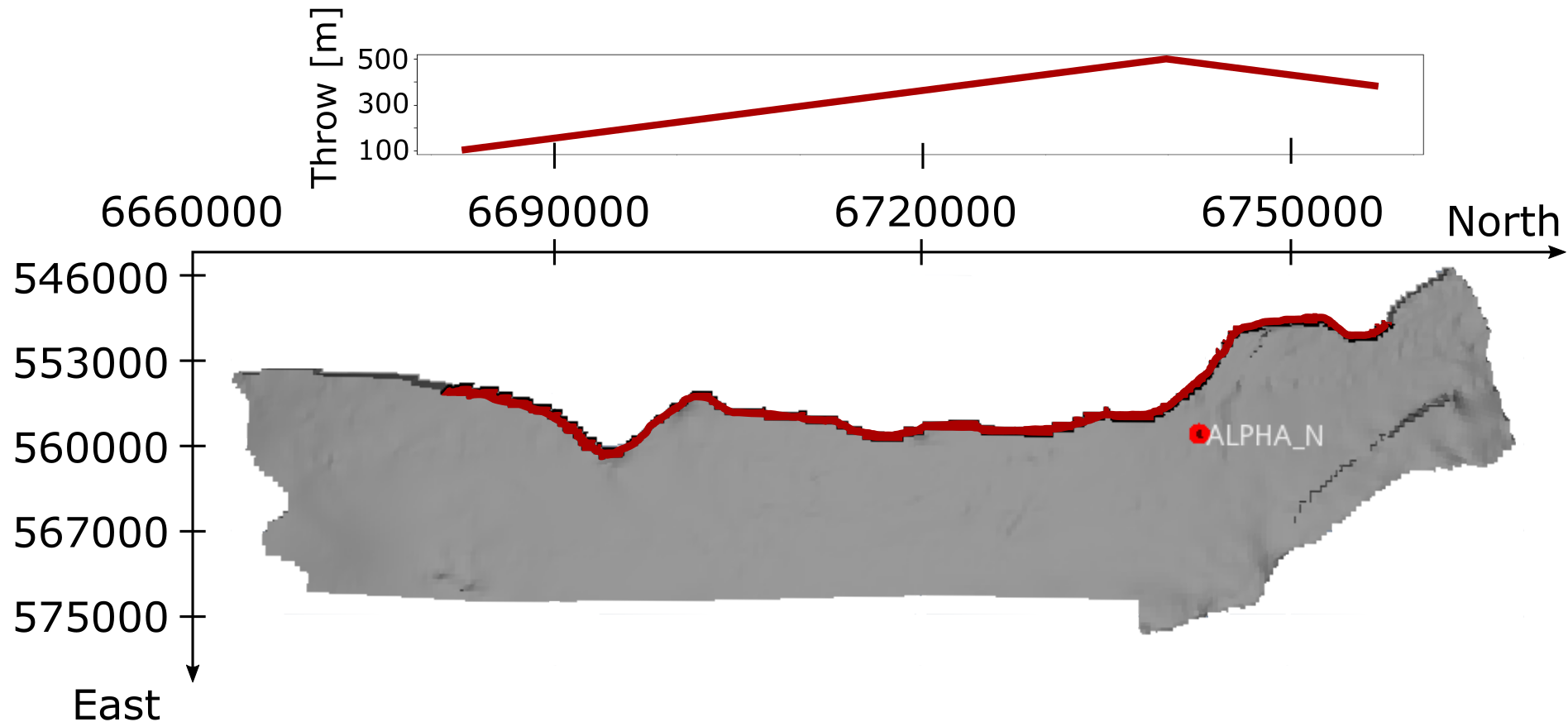}
    \caption{Top view of the Smeaheia model with the Vette fault marked in red. The red dot marks the injector location. The plot above the Smeaheia model shows the height of the Vette fault used in this study.}
    \label{fig:smeaheia_vette}
\end{figure}

A 3-dimensional simulation model for \co injection in the Smeaheia formation is openly available through the CO2{D}ata{S}hare project~\cite{furre2021}. 
The model includes the full description of simulation data, including geological, fluid and petrophysical data. In addition, an ECLIPSE model of the storage site is available, which we simulate using the open-source simulator OPM Flow \cite{rasmussen2019open}.
The formation is located at an approximate depth of 1 500 m, with the injection site placed at roughly 1 500 m depth and 3 000 m from the fault Vette. Figure~\ref{fig:smeaheia_vette} shows a top-down view of the reservoir.
The initial pressure is taken as hydrostatic.
For further information on parameters, we refer to the description of the dataset \cite{furre2021}.

In the simulations we use the dynamic ECLIPSE model from the pre-feasibility phase as it is given in the CO2DataShare project, with the following modifications to simplify the simulations.
Since our analysis of the deformation bands only consider single-phase flow, we consider only 2 years of \co injection with a constant rate of 2~202~000~Sm$^3$/day (Sm$^3$ is the volume of gas at surface condition in cubic meters) to ensure that \co does not reach the near-fault region.
Moreover, the production wells included in the original model, used to emulate the effect of depletion due to hydrocarbon production from the nearby Troll field, are shut.

In addition to the slight modification of the schedule above, the simulation model is amended to include deformation bands in the damage zone of the Vette fault, as well as possible leakage through the fault. 
The main focus of this study is on the effect of the deformation bands, and not on flow through the fault, thus, we set up a simplified fault leakage model using a numerical aquifer and non-neighbouring connections as explained in Section~\ref{sec:deformation_bandsfield_scale}. 
The aquifer added to the model represents flow along the fault as depicted in Figure~\ref{fig:conceptual_num_aquifers}:
\begin{figure}
    \centering
    \newcommand{\figWidth}{0.3\textwidth}
    \begin{subfigure}[b]{\figWidth}
    \def\svgwidth{\textwidth}
        \scriptsize
\begingroup%
  \makeatletter%
  \providecommand\color[2][]{%
    \errmessage{(Inkscape) Color is used for the text in Inkscape, but the package 'color.sty' is not loaded}%
    \renewcommand\color[2][]{}%
  }%
  \providecommand\transparent[1]{%
    \errmessage{(Inkscape) Transparency is used (non-zero) for the text in Inkscape, but the package 'transparent.sty' is not loaded}%
    \renewcommand\transparent[1]{}%
  }%
  \providecommand\rotatebox[2]{#2}%
  \newcommand*\fsize{\dimexpr\f@size pt\relax}%
  \newcommand*\lineheight[1]{\fontsize{\fsize}{#1\fsize}\selectfont}%
  \ifx\svgwidth\undefined%
    \setlength{\unitlength}{228.50020521bp}%
    \ifx\svgscale\undefined%
      \relax%
    \else%
      \setlength{\unitlength}{\unitlength * \real{\svgscale}}%
    \fi%
  \else%
    \setlength{\unitlength}{\svgwidth}%
  \fi%
  \global\let\svgwidth\undefined%
  \global\let\svgscale\undefined%
  \makeatother%
  \begin{picture}(1,0.84424127)%
    \lineheight{1}%
    \setlength\tabcolsep{0pt}%
    \put(0,0){\includegraphics[width=\unitlength,page=1]{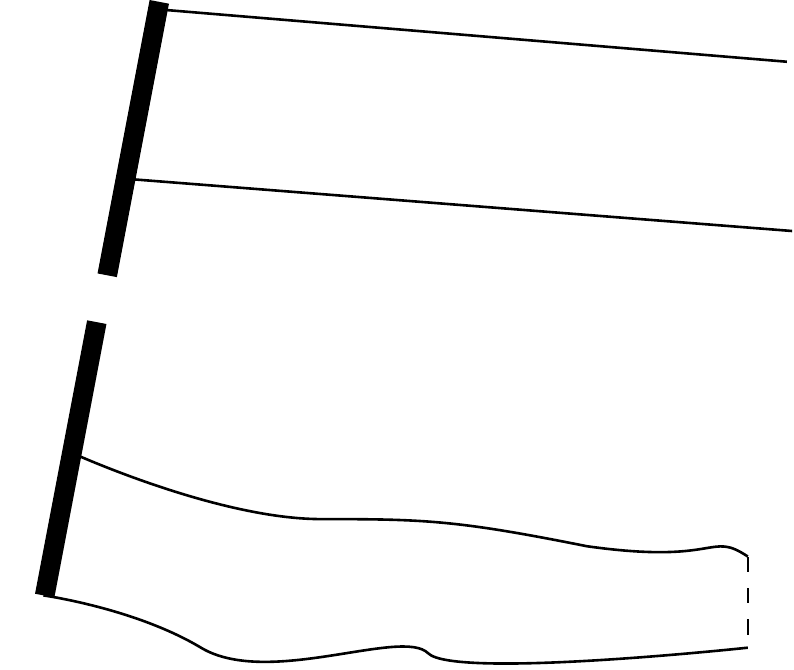}}%
    \put(0.61962855,0.646113){\makebox(0,0)[lt]{\lineheight{1.25}\smash{\begin{tabular}[t]{l}Aquifer\end{tabular}}}}%
    \put(0,0){\includegraphics[width=\unitlength,page=2]{numerical_aquifers.pdf}}%
    \put(0.21155968,0.43382978){\makebox(0,0)[lt]{\lineheight{1.25}\smash{\begin{tabular}[t]{l}$L$\end{tabular}}}}%
    \put(0,0){\includegraphics[width=\unitlength,page=3]{numerical_aquifers.pdf}}%
    \put(0.07453297,0.44189438){\rotatebox{79.139657}{\makebox(0,0)[lt]{\lineheight{1.25}\smash{\begin{tabular}[t]{l}Vette\end{tabular}}}}}%
    \put(0,0){\includegraphics[width=\unitlength,page=4]{numerical_aquifers.pdf}}%
    \put(0.36169748,0.08566198){\makebox(0,0)[lt]{\lineheight{1.25}\smash{\begin{tabular}[t]{l}Reservoir model\end{tabular}}}}%
    \put(-0.001312,0.03579709){\rotatebox{-15.753838}{\makebox(0,0)[lt]{\lineheight{1.25}\smash{\begin{tabular}[t]{l}$A_f$\end{tabular}}}}}%
  \end{picture}%
\endgroup%

    \end{subfigure}
    \caption{Conceptual model of reservoir model. \co is injected in the reservoir model that is bounded by the Vette fault. The Vette fault is connected to an aquifer located above the injection reservoir. \label{fig:conceptual_num_aquifers}}
\end{figure}
The aquifer is located above the reservoir and the pressure is initially hydrostatic.
The aquifer is connected through non-neighbouring connections (NNCs) to all cells in the reservoir that are attached to the Vette fault, and we vary the transmissibility of the connection between different simulations. The porosity of the aquifer is set sufficiently high to avoid pressure build-up that would affect the flow through the fault. 
With this setup the aquifer mimics open boundary conditions where the flow is controlled by the fault transmissibility. 
Further technical details on how to setup the simulation model with the numerical aquifers are described in the appendix.

The upscaled permeability of the deformation bands depends on four parameters; band length, band rotation, band density, and scaled permeability ratio. The band length and band rotation is fixed in the simulations to $l=2$ m and $\sigma = \pi/12$. The band density is calculated from the logarithmic function given by Equation~\eqref{eq:fault_density}, and only depends on the fault throw. The throw of the Vette fault obtained from the seismic imaging shows local variations~\cite{mulrooney2020}, but for simplicity we approximate the throw using a piecewise linear profile (see Figure~\ref{fig:smeaheia_vette}), with a maximum throw of 500 m.
The width of the damage zone is calculated from the throw according to Equation~\eqref{eq:widthFromThrow}. 

To gauge the impact of the deformation bands, we set up a suite of simulations where we vary the fault permeability and the scaled permeability ratio $K_b/(aK_m)$. For reference, we also perform simulations with equal fault transmissibility, but without deformation bands. 

\subsection{Results}
\label{sec:sim_res}
\begin{figure}
    \centering
    \includegraphics[width=0.49\textwidth]{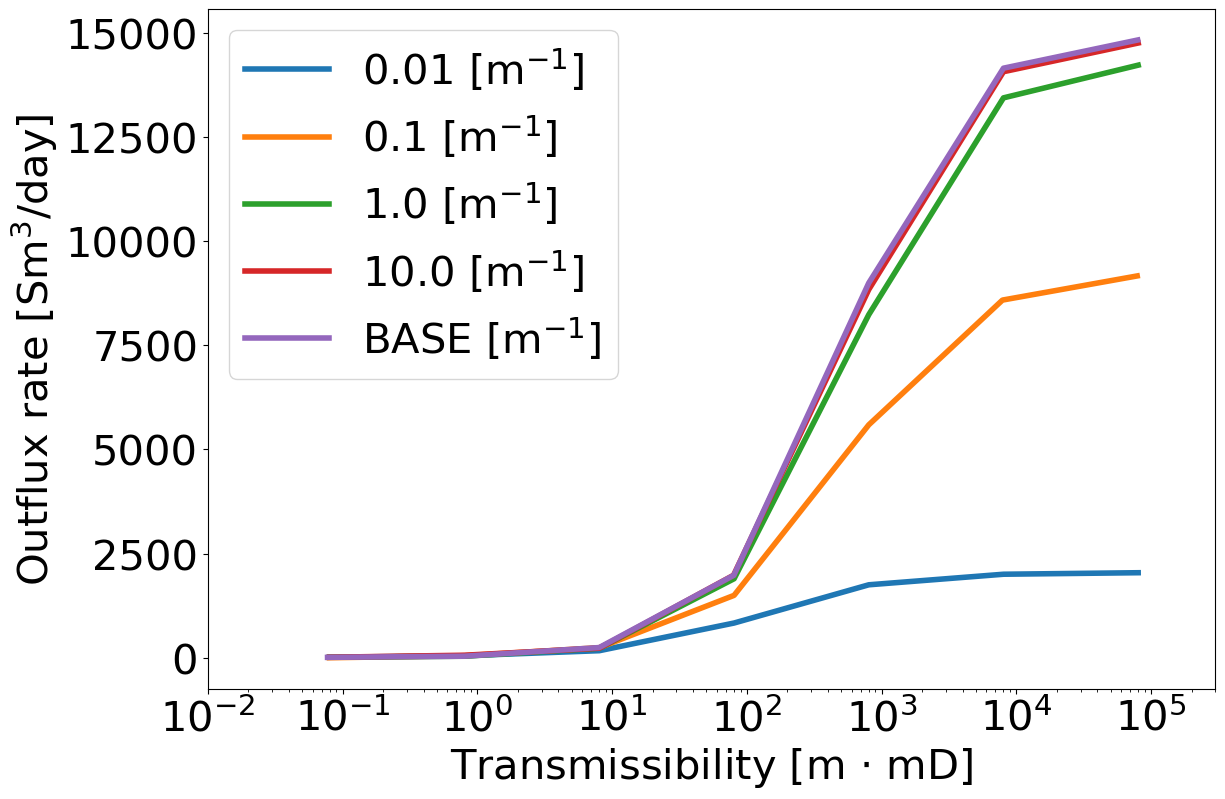}
    \caption{Outflux from the reservoir to the aquifer as function of fault transmissibility $T_{aq}$ (see Equation~\eqref{eq:num_aqf_trans}). Each line correspond to a different value of the scaled permeability ratio ($K_b/(aK_m)$). BASE case is without the effect of the deformation bands.}
    \label{fig:outflux_smeaheia}
\end{figure}

The flow rate into the aquifer after the two year injection is shown in Figure~\ref{fig:outflux_smeaheia}. The figure shows the result of simulations with different scaled permeability ratios and fault transmissibilities. The transmissibilities can be related to physical quantities by Equation~\eqref{eq:num_aqf_trans}, see also Apendix~\ref{sec:apendix}.

From the graphs in the figure we identify three regimes. In the first regime, the fault is blocking (any significant) flow out of the aquifer and the simulations show no effects on the outflux rates by adding deformation bands. In this case the transmissibility to the aquifer can be approximated by
\begin{equation*}
    T_{r-aq} = \frac{M_{db} T_{r} T_{aq}}{M_{db} T_{r} + T_{aq}} \approx T_{aq}, \qquad \text{when } T_{aq} \ll M_{db}T_r.
\end{equation*}
Thus, when the fault is sealing, then adding an extra layer of low permeable deformation bands does not influence the outflux rate.

For simulations with higher fault transmissibility, a second regime appears where the fault transmissibility, $T_{aq}$, balances the reservoir and damage zone transmissibility $M T_r$. Thus, the outflux depends on both the fault transmissibility $T_{aq}$ and the scaled permeability ratio $K_b/(aK_m)$. In this second regime, the outflux is reduced by up to an order of magnitude by the deformation bands. 

In the third regime, the fault is more transmissible than the reservoir, and the outflux rates do not depend on the fault transmissibility. 
While this is not a geologically realistic scenario, it is included to give a general analysis of the system. The outflux rate in this case is only depending on the factor $M_{db}T_r$:
\begin{equation*}
    T_{r-aq} = \frac{M_{db} T_{r} T_{aq}}{M_{db} T_{r} + T_{aq}} \approx M_{db} T_{r}, \qquad \text{when } T_{aq} \gg M_{db}T_r.
\end{equation*}
In Figure~\ref{fig:outflux_smeaheia}, this can be seen as the plateau of the outflux rates for fault transmissibilities $\gtrapprox 10^4$ m$\cdot$mD. 

The results show that the deformation bands are able to significantly decrease the outflux flow, but only under two conditions; small scaled permeability ratio ($K_b/(aK_m)\le 0.1$ m$^{-1}$) and high fault transmissibility ($T_{aq}\ge 10^1$ m$\cdot$mD). For the smallest scaled permeability ratio included in the study ($K_b/(aK_m)$ = 0.01 m$^{-1}$) the outflux is reduced by approximately an order of magnitude compared to the Base case for the moderate to high fault transmissibilities. The deformation bands, thus, may indeed act as an extra guard around the faults to prevent or restrict flow in the fault zone.

Equally important, the example illustrates that the effect of deformation bands on fluid flow through faults can be included in field-scale simulations with minimal computational cost and only minor modifications of standard model setups.
Variations in geological properties, for instance to incorporate geological uncertainty, can be incorporated  non-invasively.

\section{Discussion}
\label{sec:discussion}
The numerical results in Sections~\ref{sec:res_homogen} and~\ref{sec:res_fault} show that applying the harmonic average uncritically to upscale the effective permeability of deformation bands may over-estimate the permeability reduction due to the deformation bands by several orders of magnitude. Instead, applying a layered conceptual model where a weighted average of the harmonic and arithmetic average is used to calculate the effective permeability is a better approach for disconnected networks. The main result of this paper is an analytical approximation of how the weights in the layered model can be calculated by considering a minimal set of parameters, and a method to include the model in field scale simulations. To apply the analytical approximation the length of the deformation bands, the rotation of the deformation bands and the deformation band density must to be specified.

The two most important properties of the deformation bands that influence flow are the scaled permeability ratio and the connectiveness of the deformation band network. Previous studies of deformation bands have shown that the permeability ratio between the host rock and deformation bands must be 3-4 orders of magnitude for the deformation bands to have an effect on the flow~\cite{matthai1998,sternlof2004,fossen2007}, while for certain cases the deformation bands may influence flow for ratios as small as one order of magnitude~\cite{rotevatn2017}. It is important to note, however, that it is not only sufficient to study the permeability contrast between the deformation bands and rock matrix to draw conclusions on whether the deformation bands have a significant impact on flow. For Darcy flow the aperture of the deformation bands is equally important, and we include the aperture in what we call the scaled permeability ratio. In the numerical simulations in Sections~\ref{sec:res_homogen} and~\ref{sec:res_fault} we observe that in most cases the scaled permeability ratio must be $10^0$ m$^{-1}$ or lower for the deformation bands to influence the effective permeability significantly. This is consistent with the previous studies; if the aperture of the deformation bands is $1$ mm, the scaled permeability ratio of $K_b/(aK_m) = 10^0$ m$^{-1}$ corresponds to a permeability ratio between the rock matrix and deformation bands of 3 orders of magnitude.

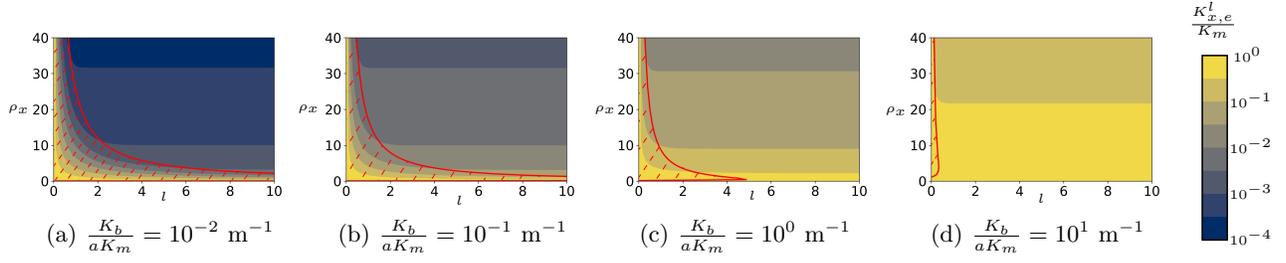
\begin{figure}
    \centering
    \newcommand{\figWidth}{0.22\textwidth}
    \begin{subfigure}[b]{\figWidth}
    \def\svgwidth{\textwidth}
    \tiny
\begingroup%
  \makeatletter%
  \providecommand\color[2][]{%
    \errmessage{(Inkscape) Color is used for the text in Inkscape, but the package 'color.sty' is not loaded}%
    \renewcommand\color[2][]{}%
  }%
  \providecommand\transparent[1]{%
    \errmessage{(Inkscape) Transparency is used (non-zero) for the text in Inkscape, but the package 'transparent.sty' is not loaded}%
    \renewcommand\transparent[1]{}%
  }%
  \providecommand\rotatebox[2]{#2}%
  \newcommand*\fsize{\dimexpr\f@size pt\relax}%
  \newcommand*\lineheight[1]{\fontsize{\fsize}{#1\fsize}\selectfont}%
  \ifx\svgwidth\undefined%
    \setlength{\unitlength}{432bp}%
    \ifx\svgscale\undefined%
      \relax%
    \else%
      \setlength{\unitlength}{\unitlength * \real{\svgscale}}%
    \fi%
  \else%
    \setlength{\unitlength}{\svgwidth}%
  \fi%
  \global\let\svgwidth\undefined%
  \global\let\svgscale\undefined%
  \makeatother%
  \begin{picture}(1,0.66666667)%
    \lineheight{1}%
    \setlength\tabcolsep{0pt}%
    \put(0,0){\includegraphics[width=\unitlength,page=1]{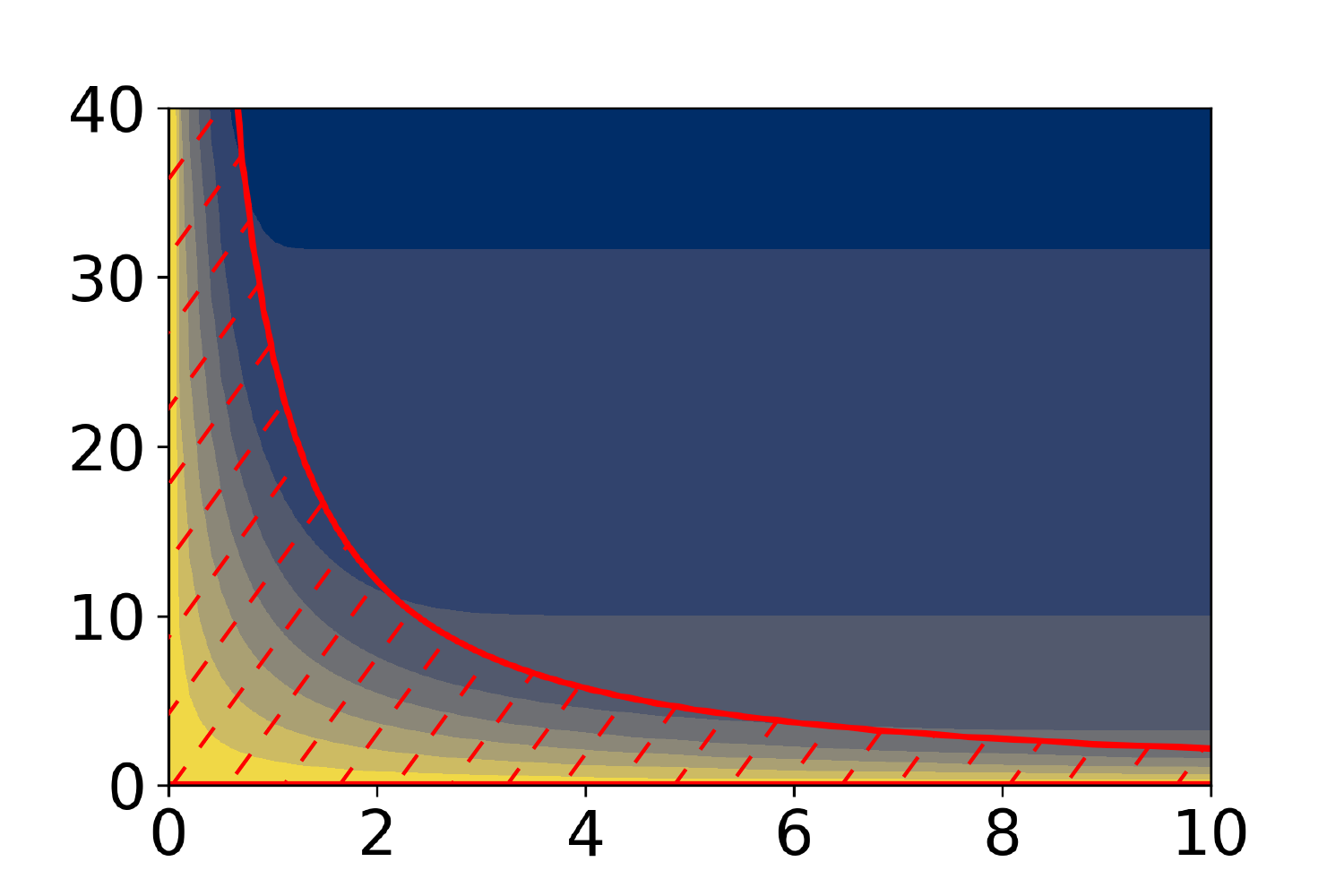}}%
    \put(0.51308372,0.01062801){\makebox(0,0)[t]{\lineheight{1.25}\smash{\begin{tabular}[t]{c}$l$\end{tabular}}}}%
    \put(0.04011582,0.32996681){\makebox(0,0)[rt]{\lineheight{1.25}\smash{\begin{tabular}[t]{r}$\rho_x$\end{tabular}}}}%
  \end{picture}%
\endgroup%

    \caption{$\frac{K_b}{aK_m} = 10^{-2}$ m$^{-1}$}
    \end{subfigure}
    \begin{subfigure}[b]{\figWidth}
    \def\svgwidth{\textwidth}
    \tiny
\begingroup%
  \makeatletter%
  \providecommand\color[2][]{%
    \errmessage{(Inkscape) Color is used for the text in Inkscape, but the package 'color.sty' is not loaded}%
    \renewcommand\color[2][]{}%
  }%
  \providecommand\transparent[1]{%
    \errmessage{(Inkscape) Transparency is used (non-zero) for the text in Inkscape, but the package 'transparent.sty' is not loaded}%
    \renewcommand\transparent[1]{}%
  }%
  \providecommand\rotatebox[2]{#2}%
  \newcommand*\fsize{\dimexpr\f@size pt\relax}%
  \newcommand*\lineheight[1]{\fontsize{\fsize}{#1\fsize}\selectfont}%
  \ifx\svgwidth\undefined%
    \setlength{\unitlength}{432bp}%
    \ifx\svgscale\undefined%
      \relax%
    \else%
      \setlength{\unitlength}{\unitlength * \real{\svgscale}}%
    \fi%
  \else%
    \setlength{\unitlength}{\svgwidth}%
  \fi%
  \global\let\svgwidth\undefined%
  \global\let\svgscale\undefined%
  \makeatother%
  \begin{picture}(1,0.66666667)%
    \lineheight{1}%
    \setlength\tabcolsep{0pt}%
    \put(0,0){\includegraphics[width=\unitlength,page=1]{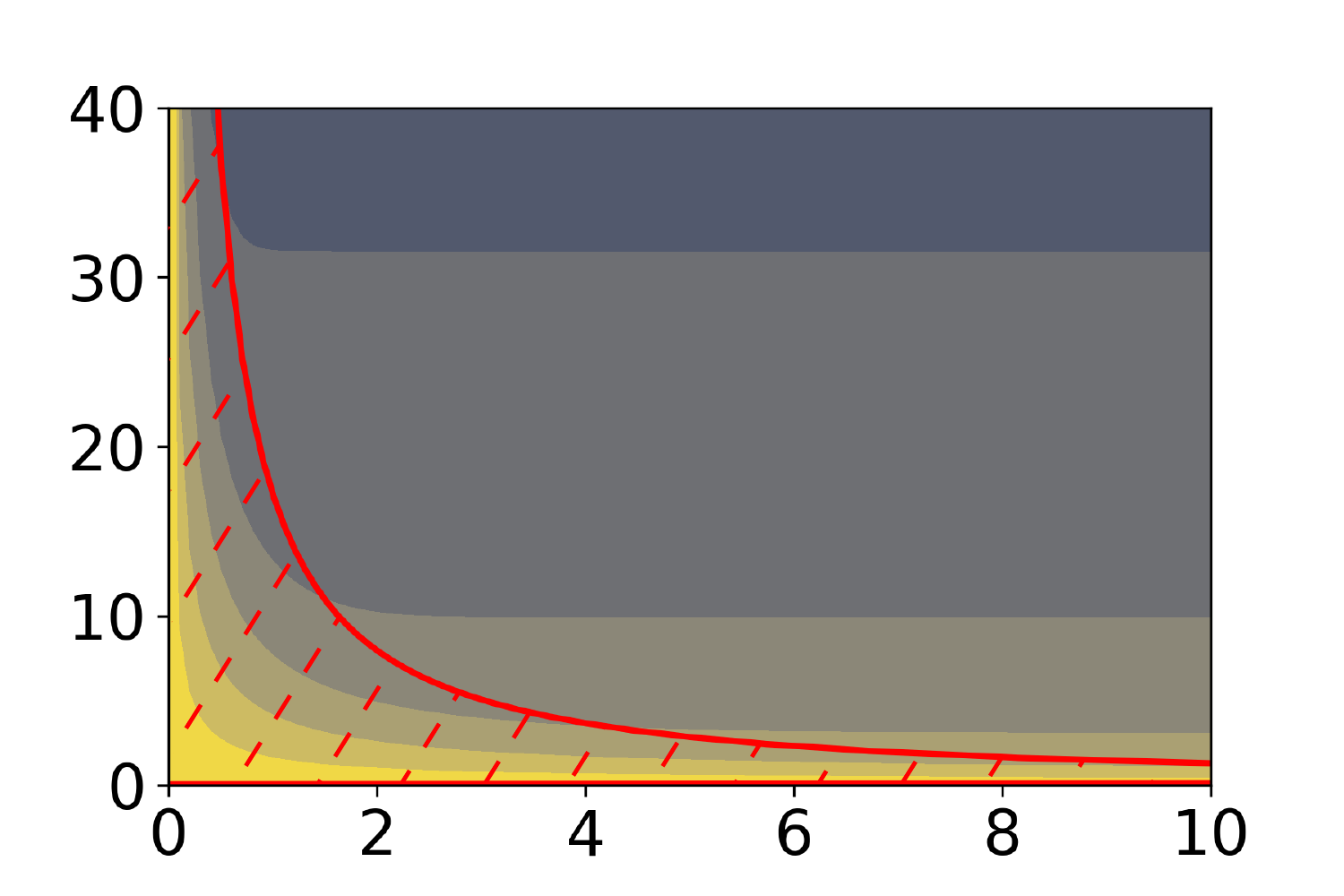}}%
    \put(0.03069366,0.33204773){\makebox(0,0)[rt]{\lineheight{1.25}\smash{\begin{tabular}[t]{r}$\rho_x$\end{tabular}}}}%
    \put(0.53505903,-0.00082201){\makebox(0,0)[rt]{\lineheight{1.25}\smash{\begin{tabular}[t]{r}$l$\end{tabular}}}}%
  \end{picture}%
\endgroup%

    \caption{$\frac{K_b}{aK_m} = 10^{-1}$ m$^{-1}$}
    \end{subfigure}
    \begin{subfigure}[b]{\figWidth}
    \def\svgwidth{\textwidth}
    \tiny
\begingroup%
  \makeatletter%
  \providecommand\color[2][]{%
    \errmessage{(Inkscape) Color is used for the text in Inkscape, but the package 'color.sty' is not loaded}%
    \renewcommand\color[2][]{}%
  }%
  \providecommand\transparent[1]{%
    \errmessage{(Inkscape) Transparency is used (non-zero) for the text in Inkscape, but the package 'transparent.sty' is not loaded}%
    \renewcommand\transparent[1]{}%
  }%
  \providecommand\rotatebox[2]{#2}%
  \newcommand*\fsize{\dimexpr\f@size pt\relax}%
  \newcommand*\lineheight[1]{\fontsize{\fsize}{#1\fsize}\selectfont}%
  \ifx\svgwidth\undefined%
    \setlength{\unitlength}{432bp}%
    \ifx\svgscale\undefined%
      \relax%
    \else%
      \setlength{\unitlength}{\unitlength * \real{\svgscale}}%
    \fi%
  \else%
    \setlength{\unitlength}{\svgwidth}%
  \fi%
  \global\let\svgwidth\undefined%
  \global\let\svgscale\undefined%
  \makeatother%
  \begin{picture}(1,0.66666667)%
    \lineheight{1}%
    \setlength\tabcolsep{0pt}%
    \put(0,0){\includegraphics[width=\unitlength,page=1]{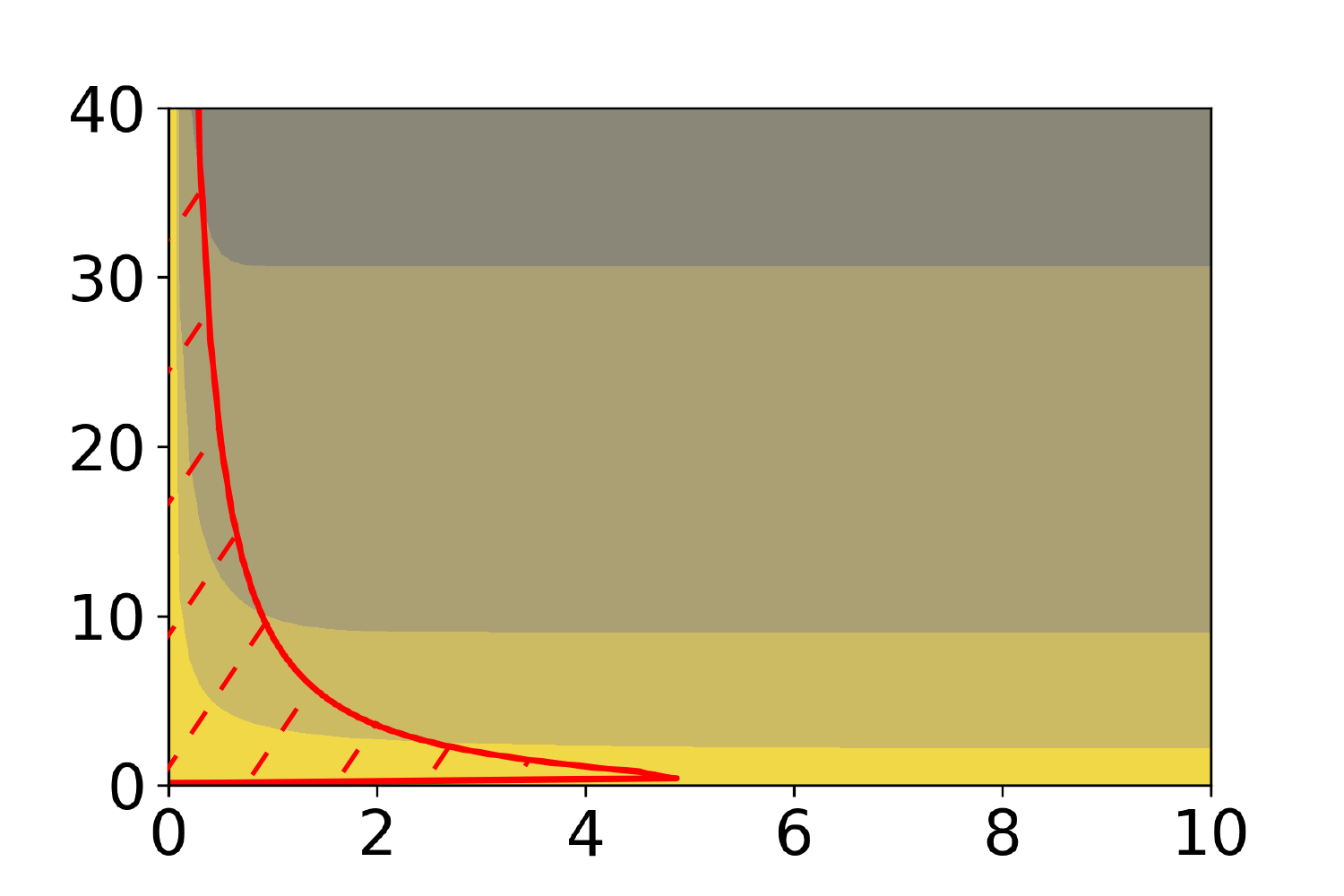}}%
    \put(0.04459056,0.32878881){\makebox(0,0)[rt]{\lineheight{1.25}\smash{\begin{tabular}[t]{r}$\rho_x$\end{tabular}}}}%
    \put(0.53671498,0.01458931){\makebox(0,0)[rt]{\lineheight{1.25}\smash{\begin{tabular}[t]{r}$l$\end{tabular}}}}%
  \end{picture}%
\endgroup%

    \caption{$\frac{K_b}{aK_m} = 10^{0}$ m$^{-1}$}
    \end{subfigure}
    \begin{subfigure}[b]{\figWidth}
    \def\svgwidth{\textwidth}
    \tiny
\begingroup%
  \makeatletter%
  \providecommand\color[2][]{%
    \errmessage{(Inkscape) Color is used for the text in Inkscape, but the package 'color.sty' is not loaded}%
    \renewcommand\color[2][]{}%
  }%
  \providecommand\transparent[1]{%
    \errmessage{(Inkscape) Transparency is used (non-zero) for the text in Inkscape, but the package 'transparent.sty' is not loaded}%
    \renewcommand\transparent[1]{}%
  }%
  \providecommand\rotatebox[2]{#2}%
  \newcommand*\fsize{\dimexpr\f@size pt\relax}%
  \newcommand*\lineheight[1]{\fontsize{\fsize}{#1\fsize}\selectfont}%
  \ifx\svgwidth\undefined%
    \setlength{\unitlength}{432bp}%
    \ifx\svgscale\undefined%
      \relax%
    \else%
      \setlength{\unitlength}{\unitlength * \real{\svgscale}}%
    \fi%
  \else%
    \setlength{\unitlength}{\svgwidth}%
  \fi%
  \global\let\svgwidth\undefined%
  \global\let\svgscale\undefined%
  \makeatother%
  \begin{picture}(1,0.66666667)%
    \lineheight{1}%
    \setlength\tabcolsep{0pt}%
    \put(0,0){\includegraphics[width=\unitlength,page=1]{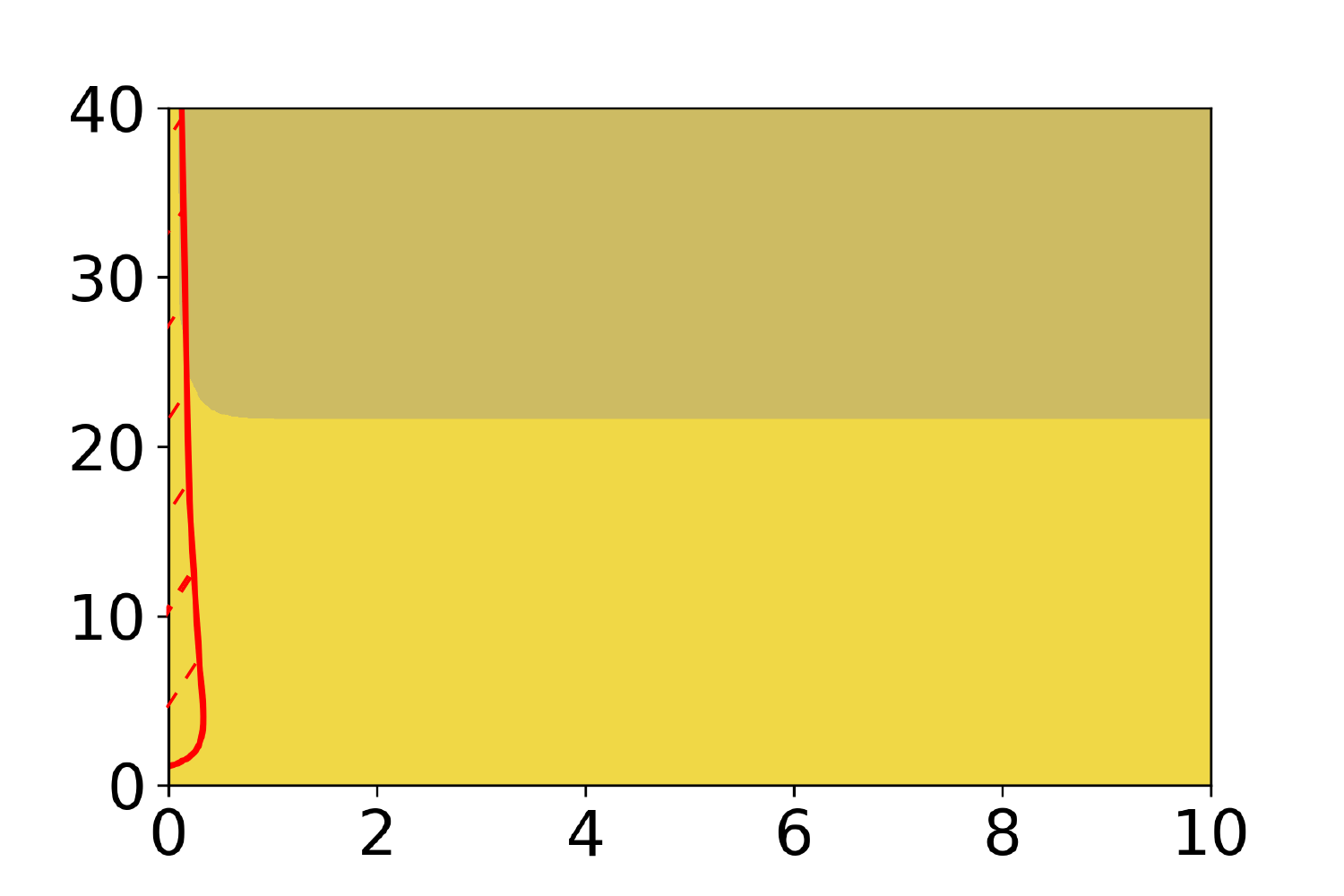}}%
    \put(0.03915906,0.32405096){\makebox(0,0)[rt]{\lineheight{1.25}\smash{\begin{tabular}[t]{r}$\rho_x$\end{tabular}}}}%
    \put(0.52360222,0.00409905){\makebox(0,0)[rt]{\lineheight{1.25}\smash{\begin{tabular}[t]{r}$l$\end{tabular}}}}%
  \end{picture}%
\endgroup%

    \caption{$\frac{K_b}{aK_m} = 10^{1}$ m$^{-1}$}
    \end{subfigure}
    \begin{subfigure}[b]{0.04\textwidth}
    \def\svgwidth{\textwidth}
    \tiny
\begingroup%
  \makeatletter%
  \providecommand\color[2][]{%
    \errmessage{(Inkscape) Color is used for the text in Inkscape, but the package 'color.sty' is not loaded}%
    \renewcommand\color[2][]{}%
  }%
  \providecommand\transparent[1]{%
    \errmessage{(Inkscape) Transparency is used (non-zero) for the text in Inkscape, but the package 'transparent.sty' is not loaded}%
    \renewcommand\transparent[1]{}%
  }%
  \providecommand\rotatebox[2]{#2}%
  \newcommand*\fsize{\dimexpr\f@size pt\relax}%
  \newcommand*\lineheight[1]{\fontsize{\fsize}{#1\fsize}\selectfont}%
  \ifx\svgwidth\undefined%
    \setlength{\unitlength}{27.35999966bp}%
    \ifx\svgscale\undefined%
      \relax%
    \else%
      \setlength{\unitlength}{\unitlength * \real{\svgscale}}%
    \fi%
  \else%
    \setlength{\unitlength}{\svgwidth}%
  \fi%
  \global\let\svgwidth\undefined%
  \global\let\svgscale\undefined%
  \makeatother%
  \begin{picture}(1,4.0789475)%
    \lineheight{1}%
    \setlength\tabcolsep{0pt}%
    \put(0,0){\includegraphics[width=\unitlength,page=1]{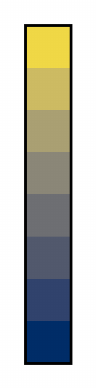}}%
    \put(0.79742755,0.1645197){\makebox(0,0)[lt]{\lineheight{1.25}\smash{\begin{tabular}[t]{l}$10^{-4}$\end{tabular}}}}%
    \put(0.79742755,1.05194775){\makebox(0,0)[lt]{\lineheight{1.25}\smash{\begin{tabular}[t]{l}$10^{-3}$\end{tabular}}}}%
    \put(0.79742755,1.93937559){\makebox(0,0)[lt]{\lineheight{1.25}\smash{\begin{tabular}[t]{l}$10^{-2}$\end{tabular}}}}%
    \put(0.79742755,2.82680385){\makebox(0,0)[lt]{\lineheight{1.25}\smash{\begin{tabular}[t]{l}$10^{-1}$\end{tabular}}}}%
    \put(0.86381687,3.71423191){\makebox(0,0)[lt]{\lineheight{1.25}\smash{\begin{tabular}[t]{l}$10^{0}$\end{tabular}}}}%
    \put(0.94171293,4.36005104){\makebox(0,0)[rt]{\lineheight{1.25}\smash{\begin{tabular}[t]{r}$\frac{K_{x,e}^l}{K_m}$\end{tabular}}}}%
  \end{picture}%
\endgroup%

    \end{subfigure}
    \caption{Contour plots of the effective permeability in the $x$-direction calculated by the using the layered model. The $x$-axis represents the band length and the $y$-axis represents the density along a scanline $\rho_x$. The region with dashed red lines indicates where using the harmonic average will over-estimates the permeability reduction due to the deformation bands by more than 10 \%.}
    \label{fig:Kxlb_vs_l_vs_rho}
\end{figure}

The network connectivity is a second property that can influence the upscaled permeability greatly. A disconnected network with paths for the fluid around the deformation bands has a smaller influence on flow than a highly connected network where the fluid must cross the deformation bands. In Section~\ref{sec:res_homogen} we can see that the upscaled permeability has two distinct regimes. When the band length or band density increases, then the upscaled permeability do approach the harmonic average. The steepest part of the permeability curves appears when the expected number of crossings (calculated from Equation~\eqref{eq:expected_number_of_cross}) of a deformation band is 2. For a lower number of crossings the network mainly consists of individual deformation bands with few intersecting bands. 
Thus, it is insufficient to use only the scaled permeability ratio of the deformation bands and the rock matrix to estimate the effective permeability. Figure~\ref{fig:Kxlb_vs_l_vs_rho} shows contour plots of the effective permeability for different scan-line densities $\rho_x$ and band lengths $l$. The red-dashed areas indicate the regions where the error of using the harmonic average is larger than 10 \%. While the harmonic average gives appropriate results for high densities and band lengths, the region where the harmonic average is not applicable is considerable, especially, for large permeability contrasts.

For deformation band networks that do not follow the statistical distributions in this paper, the quantitative values (e.g., in Figure~\ref{fig:Kxlb_vs_l_vs_rho}) should be used with care, however, we can still draw insight from simulations. If the network of interest consists of long or continuous connected deformation bands using the harmonic average to calculate the effective permeability is appropriate. On the other hand, if the network mainly consists of short individual bands with paths for the fluid to flow around the deformation bands a layered model should be used instead. This layered model will give good results provided good estimates of the areas available for flow around the deformation bands can be found.

The upscaling of the permeability in this paper is only done for single-phase flow and we do not investigate how the deformation bands affect the relative permeability and capillary pressures. Including these effects will be crucial in a two-phase scenario, and it is expected that capillary pressures should enhance the blocking effect of the deformation bands to \co due to the smaller pore size in the deformation bands~\cite{torabi2013}. Upscaling of two-phase effects is necessary in order to give quantitative results of leakage rates of \co through the damage zone of faults. The upscaled functions of the relative permeability and the capillary pressure will of cause be highly dependent on the parameter contrast between the rock matrix and the deformation bands, but it will, presumably, also be as sensitive to the geometry of the deformation bands as the intrinsic permeability has been shown to be in this paper. The upscaling of the intrinsic permeability in this paper is a first step in this direction.

The field scale simulations presented in Section \ref{sec:field_scale_sim} show that the permeability reduction from deformation bands can act as a secondary seal for a fault.
While this extra seal is of little importance for faults of low permeability, deformation bands can be of importance for faults that potentially permits fluid flow. 
Our simulations focused on fluid flow along the fault but the deformation bands will also have a partially blocking effect on fluid flow and pressure communication across the fault.

Sections \ref{sec:deformation_bandsfield_scale} and \ref{sec:field_scale_sim} also showed that the permeability reduction due to deformation bands can readily be included in field-scale simulations.
Moreover, the study showed how standard simulation models can be adapted with minimal modifications to include the interplay between flow in the storage formation and possible leakage through a fault. 
The techniques used, transmissibility multipliers and numerical aquifers that are connected to the main injection formation through non-neighbor connections, are standard in reservoir simulation and has minimal overhead in simulation setup and computational cost.
Our approach is ideally suited for uncertainty quantification applied to deformation bands, but also to stochastic representations of the flow properties of the fault.
Future work in this direction includes benchmarking of the numerical aquifer approach, extension of the methodology to handle two-phase flow, and allowing for potential increases in the fault permeability as increases in the fluid pressure leads to a reduction in the effective stress.

\section{Concluding remarks}
\label{sec:conclusion}
This paper studies how deformation bands impact fluid flow in the damage zone of faults. Two different methods for upscaling the effective permeability of deformation bands are validated against numerical simulations; a harmonic average and a novel layered model. It is found that in many cases it is necessary to consider the geometry of the deformation band network to obtain good estimates of the effective permeability. Thus, the harmonic average overestimates the permeability reduction due to the deformation bands when the deformation band network consists of short disconnected bands. On the other hand, if the deformation band network is well connected the harmonic average gives appropriate estimations of the effective permeability. The proposed layered model gives better estimates of the effective permeability, and it captures the increase in effective permeability when the network consists of short disconnected deformation bands, and reduces to the harmonic average when appropriate.

The upscaled effective permeability can be included in field-scale simulations of storage reservoirs as transmissibility multiplicators associated with computational cell faces. By studying an injection scenario in the Smeaheia storage site, it is observed that the deformation bands can cause a significant reduction of leakage rates through faults. This requires a relative high permeability contrast between the rock matrix and deformation bands (three orders of magnitude) and a deformation band aperture of at least $1$ mm. Thus, deformation bands in the damage zone may act as an extra barrier that reduce the fluid flow along or across faults.

\section{Acknowledgments}
This work was financed in part by the Norwegian Research Council grant 294719.

\section*{Appendix}
\appendix
\section{Details on how to setup the simulation model}\label{sec:apendix}
  To ensure reproducibility of the simulations some technical details on needed adjustments of the simulation model downloaded from CO2DataShare is included here. The standard approach for representing \co-brine systems in reservoir simulators that are originally designed for oil-gas systems is to denote \co and brine as gas and oil, respectively, and thereby allow for dissolution of \co into the brine.
This is the approach taken in the original model from CO2DataShare, however, it is incompatible with the use of numerical aquifers that are assumed to be filled with the water phase. One way to circumvent this issue is to represent the brine by the water phase and, thus, ignore dissolution of \co into brine. An alternative is to use the CO2STORE option available in OPM Flow which is compatible with numerical aquifers by assuming the aquifers to be filled with the phase labeled oil used for the brine. We have chosen the latter approach and all the OPM Flow simulations in this work use the CO2STORE option. Another benefit of using the CO2STORE option is that the PVT properties are internal functions in the simulator and thus computes the PVT properties based on the pressure, temperature and salinity dynamically during the simulations. To use the CO2STORE option, the CO2STORE keyword needs to be added to the downloaded ECLIPSE type model in the top section and preferably all the tabulated PVT data removed from the model as they are not used anymore. For details on the used keywords please confer the OPM Flow manual~\cite{opmFlowManual}.

In addition, the numerical aquifer as described in Section \ref{sec:sim_setup} needs to be added to the model using the AQUNUM keyword. The input values are shown in Table \ref{tab:aquifer_input}. Note the length value ($L_f$ in Equation~\eqref{eq:num_aqf_trans}) needs to be multiplied by a factor 2 as the length expected by the simulator is the length of the aquifer and thus the simulator divides the length by a factor 2 internally when calculating the transmissibility. The fault permeability is also set in this keyword. To vary the transmissibility in the simulations we fix the area $A_f$ and the length $L_f$ and vary the fault permeability $K_f$. The aquifer is connected through the Vette fault and a connection map between the cells adjacent to the Vette fault and the numerical aquifer is therefore needed (see AQUCON in the OPM Flow manual~\cite{opmFlowManual}). The downloaded model unfortunately does not explicitly include Vette fault data and some pre-processing is needed to identify the relevant cells in the model.
Finally, we include the effect of the deformation band on the fluid flow using the permeability multiplier keywords MULTX- and MULTY- for the cells adjacent to the fault.
Here X- and Y- specifies that the multiplier is only applied to compute the transmissibilities pointing out-words toward the Vette fault. Different values are used for the multipliers in each cell as its value depend on the height of the fault at the cell location. For reproducibility, the fault and aquifer data used in the simulations are given in  \url{https://github.com/OPM/opm-publications}.

\begin{table}
    \centering
   \begin{tabular}{c|c c c c c c}
     Name & Area [m$^2$] ($A_f$) & Length [m] ($2L_f$) & Depth [m] & Poro [-] & Pres [bars] & Perm [mD] ($K_f$)\\
     \hline
     Value & 40 & $2\cdot500$ & 1082.9* & 1e12 & 109.1* & $1$ -- $10^6$
    \end{tabular}
    \caption{Input to numerical aquifers. The $*$ indicates that the values are defaulted and thus computed internally in the simulator. See AQUNUM in~\cite{opmFlowManual} for details.}
    \label{tab:aquifer_input}
\end{table}

\printbibliography 

\end{document}